\documentclass[aps,prb,twocolumn,superscriptaddress]{revtex4-2}
\usepackage{amsfonts, amssymb, amsmath,amsthm,graphicx}
\usepackage{mathrsfs}
\usepackage{subfigure}
\usepackage{multirow}
\usepackage{dcolumn}
\usepackage{bm}
\usepackage{color}
\usepackage[usenames,dvipsnames]{xcolor}
\usepackage[all]{xy}

\renewcommand{\bf}[1]{\textnormal{\textbf{#1}}}
\newcommand{\BZ}{\textnormal{\text{BZ}}}
\newcommand{\tr}{\textnormal{\text{Tr}}}

\newcommand{\ket}[1]{| #1 \rangle}
\newcommand{\bra}[1]{\langle #1|}

\usepackage[colorlinks=true,allcolors=blue]{hyperref}

\begin{document}

\author{Bruno Mera}
\author{Tomoki Ozawa}

\affiliation{Advanced Institute for Materials Research (WPI-AIMR), Tohoku University, Sendai 980-8577, Japan}

\title{Singular connection approach to topological phases and resonant optical responses}

\date{\today}

\begin{abstract}
We introduce a class of singular connections as an alternative to the Berry connection for any family of quantum states defined over a parameter space. We find a natural application of the singular connection in the context of transition dipoles between two bands. We find that the shift vector is nothing but the difference between the singular connection and the connection induced from the Berry connections of involved bands; the gauge invariance of the shift vector is transparent from this expression. We show, using singular connections, that the topological invariant in two dimensions associated with optical transitions between the two bands can be computed, by means of this connection, by algebraically counting the points in the zero locus of a transition dipole matrix element of the two bands involved. It follows that this invariant provides a natural topological lower bound on the number of momenta in the Brillouin zone for which an electron cannot be excited from one Bloch band to the other by absorbing a photon.
\end{abstract}

\maketitle

%
%

%
\section{Introduction}
\label{sec: introduction}

Geometry and topology of quantum states have proven to be of fundamental importance in understanding the physical behavior of materials.
Michael Berry pointed out that the geometrical phase of quantum states defined in a parameter space under adiabatic processes is gauge invariant and can have observable meanings~\cite{berry:84}; the geometrical phase is thus now also called Berry's phase~\cite{simon:83, shapere:wilczek:89,bohm:03, chruscinski:jamiolkowski:04}. In topological phases of matter, geometry and topology of quantum states defined in momentum space plays crucial roles. The topological property of Bloch states of a particle moving in a two-dimensional space under a periodic potential is known to reflect the edge properties, giving rise to the quantized Hall conductance~\cite{niu:thouless:wu:85}. The concepts often used in physics to describe the geometry of quantum states are the Berry connection and the Berry curvature. In describing the topology of the quantum states, however, one is in principle free to choose any connection, determining the associated curvature, provided the choice is consistent with the transformation properties of quantum states under gauge transformations. The reason that the Berry connection and Berry curvature are used is that they are natural and easy to compute for given quantum states, and, furthermore, the Berry curvature has observable consequences in terms of the response of the system, such as electric polarization~\cite{resta:92,smith:vanderbilt:93}, orbital magnetism~\cite{thonhauser:ceresoli:vanderbilt:resta:05,xiao:shi:niu:05}, quantum charge pumping~\cite{thouless:83, niu:thouless:84} and the celebrated anomalous quantum Hall effect in Chern insulators~\cite{haldane:88}. See also the review article~\cite{xiao:chang:niu:10} for a detailed account of Berry phase effects on electronic properties.

Considering the breadth of influences the geometry and topology of quantum states have had, the scarcity of the variety of useful connections and associated curvatures other than Berry's is perhaps surprising. In fact, very recently, different topological and geometrical structures have been found to describe the physics of resonant optical transitions~\cite{ahn:guo:nagaosa:vishwanath:22} where a novel Riemannian geometrical structure was found to describe inter-band properties. Generalizations of the Berry connection to the higher categorical setting have also been recently considered in Refs.~\cite{palumbo:goldman:19,palumbo:21}.

In this paper, we introduce an alternative connection and associated curvature for quantum states or a pair of quantum states, which behave in a singular manner over parameter space, yet define a valid connection and curvature to describe the topology of the system. Notably, the curvature becomes a two-form with delta functions as coefficients, determining an object mathematically known as a two-current, dating back to the foundational work of Georges de Rham~\cite{deRham:12}.
To define the singular connection, we need knowledge of the quantum states as well as a global section of the associated complex line bundle. The singular connection is therefore of physical significance particularly when there is a physically motivated global section. 
We find that such singular connection and curvature appear naturally in the context of resonant optical responses describing optical transitions between a pair of quantum states belonging to different energy bands, where the transition dipoles give a physically preferred global section. In particular, we find that the shift vector---a quantity associated with the shift current~\cite{sipe:shkrebtii:2000} and characterizing the interband contribution to the nonlinear optical effect of second and higher-harmonic generation~\cite{morimoto:nagaosa:16,nagaosa:morimoto:17, qian:yu:22}---can be written as a combination of contributions from the Berry connection and the singular connection. The zeros of the transition dipoles in resonant optical response are precisely the points where the singular curvature takes infinite values (in the sense of a Dirac delta function). We discuss that whenever physically preferred sections exist, the singular connection and curvature can give physically meaningful features of the quantum system.

The organization of the paper is the following. In Sec.~\ref{sec: singular connections}, starting from reviewing the Berry's connection, we define the singular connection for quantum states $\ket{\psi(\bm{\lambda})}$ defined in a general parameter space $\bm{\lambda}=(\lambda_1,\dots,\lambda_m)$. 
In Secs.~\ref{subsec: spins in magnetic fields} and~\ref{subsec: massive Dirac model}, using simple examples, we demonstrate how the singular connection and the curvature can be calculated. Sec.~\ref{subsec: spins in magnetic fields} discusses the paradigmatic example of a spin in a magnetic field and we remark that the singularity of the curvature can be interpreted in terms of Dirac strings. Section~\ref{subsec: massive Dirac model} discusses a massive Dirac Hamiltonian defined in momentum space, and we interpret topological phase transitions in terms of intersections of the Brillouin zone with Dirac strings in an extended parameter space.
In Sec.~\ref{sec: transition dipoles}, we apply the idea of singular connections to resonant optical transitions, considering a pair of quantum states $\ket{\psi_n(\bf{k})}$ and $\ket{\psi_m(\bf{k})}$ where $n$ and $m$ are band indices of the bands involved in the optical transition. We show that singular connections and the associated curvature currents can equally be well defined in this case. We show that the shift vector is nothing but the difference between the Berry connection and a singular connection, and that the topological index for the resonant optical transitions defined in~\cite{ahn:guo:nagaosa:vishwanath:22} is precisely the Chern number, which counts the number of zeros of the transition dipoles and can be calculated from the singular curvature. Sections~\ref{subsec: transition dipoles in the Bloch sphere} and~\ref{subsec: Transition dipoles in the massive Dirac model} examines the transition dipoles from the viewpoint of singular connection theory for the examples of the Bloch sphere Hamiltonian and the massive Dirac model, respectively. Finally, Sec.~\ref{sec: conclusion} concludes the paper. Mathematical details, which are omitted in the main text, can be found in Appendix.

\section{Singular connections}
\label{sec: singular connections}
In this section, we first review the Berry connection and the Berry curvature, and then introduce the singular connection and the associated curvature.

We consider a standard setup of a quantum state $\ket{\psi(\bm{\lambda})} \in \mathbb{C}^{n+1}$ varying smoothly with some parameters $\bm{\lambda}=(\lambda_1,\dots,\lambda_m)$. More formally, since quantum states are only defined up to multiplication by a nonvanishing scalar what we are looking at is a smooth map $f:M\to\mathbb{C}P^{n}$ where $M$ is the smooth manifold of dimension $m$ and $\mathbb{C}P^{n}$ is the complex projective space consisting of one-dimensional linear subspaces of $\mathbb{C}^{n+1}$.  This map $f$ assigns, locally in a chart on $M$ where $\bm{\lambda}$ defines local coordinates, the vector $\ket{\psi(\bm{\lambda})}$ up to scale. Associated to the map there is a \emph{line bundle} $\mathscr{L}\to M$ whose fiber at $\bm{\lambda}$ is the vector subspace of $\mathbb{C}^{n+1}$ defined by $f(\bm{\lambda})$ or, in other words, the line through $\ket{\psi(\bm{\lambda})}$. A {\it section} $s$ of the line bundle $\mathscr{L}$ is a map $s: \bm{\lambda} \mapsto s(\bm{\lambda})$, where $\bm{\lambda} \in M$ and $s(\bm{\lambda})$ is in the fiber of $\mathscr{L}$ over $\bm{\lambda}$. A section defined over the entire $M$ is called a {\it global section}, whereas a section defined only on an open subset of $M$ is called a {\it local section}. A particular choice of a local representation $\ket{\psi(\bm{\lambda})}$ of the map $f$ valid in some open subset of $M$ gives a local section of the line bundle $\mathscr{L} \to M$. When the line bundle $\mathscr{L} \to M$ is topologically nontrivial, the section $\ket{\psi(\bm{\lambda})}$ cannot be taken smooth and nonzero over the entire $M$. As we will see later, however, it is possible to find a global section by allowing the sections to vanish somewhere in $M$. Such a global section will play a key role in defining the singular connection.

In Michael Berry's original work, Ref.~\cite{berry:84}, a \emph{natural} connection, compatible with the adiabatic evolution of quantum states, was introduced in the line bundle $\mathscr{L}$ which we now refer to as \emph{the Berry connection.} We proceed to explain this connection and why we say it is natural. The line bundle $\mathscr{L}$ is a subbundle of the trivial rank $n+1$ vector bundle $M\times \mathbb{C}^{n+1}$ because all of these quantum states live, by assumption, in the same Hilbert space $\mathbb{C}^{n+1}$.
The Berry connection is defined, in local coordinates, by the following one-form
\begin{align}
A(\bm{\lambda})=\frac{\bra{\psi(\bm{\lambda})}d\ket{\psi(\bm{\lambda})}}{\bra{\psi(\bm{\lambda})}\psi(\bm{\lambda})\rangle}=    \sum_{i=1}^{m}\frac{\bra{\psi(\bm{\lambda})}\frac{\partial}{\partial \lambda_i}\ket{\psi(\bm{\lambda})}}{\bra{\psi(\bm{\lambda})}\psi(\bm{\lambda})\rangle} d\lambda_i.
\end{align}
It depends on the choice of gauge, i.e., the choice of the local representation $\ket{\psi(\bm{\lambda})}$ of $f$. As we change the local representation of the map $f$, $\ket{\psi(\bm{\lambda})}\to g(\bm{\lambda}) \ket{\psi(\bm{\lambda})}$ where $g(\bm{\lambda})$ is a local gauge transformation with values in the group of nonvanishing complex numbers, we have
\begin{align}
A(\bm{\lambda})\to A(\bm{\lambda}) +g(\bm{\lambda})^{-1}dg(\bm{\lambda}),
\label{eq: gauge transformations}
\end{align}
which is the desired transformation property of the connection one-form under gauge transformations.
The naturality of this connection comes from the fact that it uses no additional data other than the one already present in the map $f$ and in the line bundle $\mathscr{L}$ derived from it---namely, it uses the fact that $\mathscr{L}\subset M\times \mathbb{C}^{n+1}$ and the canonical Hermitian structure on $\mathbb{C}^{n+1}$. It is not surprising, as pointed out by Barry Simon in Ref.~\cite{simon:83}, that this connection and its curvature arise in many settings in quantum mechanics, such as the adiabatic theorem and associated abiabatic responses, simply because quantum mechanics is defined with reference to a Hilbert space which contains the Hermitian inner product structure in its definition. Of particular interest is the differential two-form called \emph{the Berry curvature}
\begin{align}
\frac{iF}{2\pi} =\frac{i dA}{2\pi},
\end{align}
because it describes the first Chern class of $\mathscr{L}$, a \emph{characteristic class}, having the property that when integrated along a closed two-dimensional submanifold of $M$ it gives an integer topological invariant known as the first Chern number.

It is a mathematical fact that on a line bundles, such as $\mathscr{L}\to M$, one can introduce many different connections. A connection being specified by local one-forms, defined on local patches on $M$, with the property that under gauge transformations they change as Eq.~\eqref{eq: gauge transformations}. Suppose now we have two connections $\nabla$ and $\nabla'$ locally described by gauge fields $A$ and $A'$. Then, its difference $\delta=\nabla -\nabla'$ is locally determined by $A-A'$ which is then gauge invariant and hence determines a globally defined one-form. It follows that the difference of curvatures $\frac{iF}{2\pi}-\frac{iF'}{2\pi}=\frac{id\delta}{2\pi}$ is an exact two-form and hence the first Chern class does not depend on the connection chosen. Note that this is true in general---characteristic classes do not depend on the choice of connection. One may then be tempted to look for other connections satisfying convenient properties. We will now introduce a class of such connections, which are singular in nature, but nevertheless perfectly valid, and have a useful localization property for their curvature.

Unlike the Berry connection, which could be uniquely defined just from the information of the line bundle $\mathscr{L}\subset M\times \mathbb{C}^{n+1}$, in order to define the singular connection, we need to assume that a global section $s$ of the line bundle is given.
A global section $s$ can be written locally in terms of the local representation $|\psi (\bm{\lambda})\rangle$ in the fiber at $\bm{\lambda}$ as $s(\bm{\lambda}) = a(\bm{\lambda}) |\psi (\bm{\lambda})\rangle$. The coefficient $a(\bm{\lambda})$ is a smooth complex function over the domain where the local representation $|\psi (\bm{\lambda})\rangle$ is defined. The local representations of the section can be patched together to give rise to a global section; upon a gauge transformation of the local representation $|\psi (\bm{\lambda})\rangle \to g(\bm{\lambda})|\psi (\bm{\lambda})\rangle$, the coefficient transforms as $a(\bm{\lambda}) \to g(\bm{\lambda})^{-1} a(\bm{\lambda})$.
For a given line bundle there are many global sections.
As we shall see, if we choose a different global section $s$, the resulting singular connection would be different; any singular connection defined from a given (generic) section would be a valid connection when discussing the topology of the system (see Appendix~\ref{sec:singular connections on line bundles associated with generic sections} for mathematical details on the section being generic). When there is a physical motivation to choose a particular section, the resulting geometrical structure of the singular connection and curvature can also become physically relevant.

In terms of a chosen global generic section $s$, we define a singular connection via \footnote{One can also check that the Berry connection can be defined in terms of a section via $\nabla s = Pds$, where $P$ is an orthogonal projection operator whose local coordinate representation is $P(\bm{\lambda})=\frac{\ket{\psi(\bm{\lambda})}\bra{\psi(\bm{\lambda})}}{\bra{\psi(\bm{\lambda})}\psi(\bm{\lambda})\rangle}$, which projects elements of $\mathbb{C}^{n+1}$ onto the fiber of $\mathscr{L}$ over $\bm{\lambda}$.}
\begin{align}
\nabla s \equiv 0.
\end{align}
Using the local representation $s(\bm{\lambda})=a(\bm{\lambda})\ket{\psi(\bm{\lambda})}$, we find that,
\begin{align}
    0 = \nabla s(\bm{\lambda}) = da(\bm{\lambda}) |\psi (\bm{\lambda})\rangle + a(\bm{\lambda}) \nabla |\psi (\bm{\lambda})\rangle.
\end{align}
In the gauge specified by $\ket{\psi(\bm{\lambda})}$, the coordinate representation of a connection is determined by how the connection acts on $\ket{\psi(\bm{\lambda})}$. Since
\begin{align}
\nabla \ket{\psi(\bm{\lambda})} = -\frac{da(\bm{\lambda})}{a(\bm{\lambda})}\ket{\psi(\bm{\lambda})},
\end{align}

the local representation of the singular connection is 
\begin{align}
A^{\text{sing}}(\bm{\lambda})= -\frac{d a(\bm{\lambda})}{a(\bm{\lambda})} = -d\ln a(\bm{\lambda}),    
\end{align}
where we wrote the superscript ``$\text{sing}$'' because the connection is singular whenever $a(\bm{\lambda})=0$, i.e., whenever the section $s$ vanishes. If we introduce local polar coordinates $a(\bm{\lambda})=r(\bm{\lambda}) e^{i\theta(\bm{\lambda})}$, we see that
\begin{align}
-\frac{d a(\bm{\lambda})}{a(\bm{\lambda})}= -\frac{dr(\bm{\lambda})}{r(\bm{\lambda})} -i d\theta(\bm{\lambda}).
\end{align}
If we were to take the exterior derivative of this equation to obtain the local form of the curvature, we would have a problem if we have a zero of $s$. The coordinate $r(\bm{\lambda})$ can be extended throughout smoothly without problem and, because of this, it turns out that we can ignore the first term for the purpose of the evaluation of $-d\left( da/a\right)$, which would naively be zero provided $\ln a(\bm{\lambda})$ was a smooth function. The true issue comes from $\theta(\bm{\lambda})$, which is not only multi-valued, if we remove the zero locus of $s$, but also, more importantly, it is not defined there. A physicist friendly approach to this problem---not completely rigorous---is as follows. In the plane defined by polar coordinates $\bf{r}=(r,\theta)$, we know that $\ln(r)$ determines the Green's function for the two-dimensional Laplacian: 
\begin{align}
\nabla^2 \ln (r)=2\pi\delta^2(\bf{r}).    
\end{align}
The standard argument for showing this result is to integrate $\nabla^2\ln (r)$ on an arbitrary disk $D_{R}$ of radius $R$ surrounding the origin, with boundary the circle $S_R=\partial D_R$, and to use a version of Stokes' theorem,
\begin{align}
\int_{D_R}\nabla^2 \ln r\;  d^2x=\int_{S_R}\left(\frac{\bf{r}}{r}\cdot \nabla \ln r \right)Rd\theta =\int_{S_R}d\theta =2\pi.
\end{align}
Additionally, because Stokes' theorem tells us $\int_{S_R}\eta=\int_{D_R}d\eta$ for one-forms $\eta$, this in turn tells us that 
\begin{align}
dd\theta =2\pi \delta^2(\bf{r})dx_1\wedge dx_2.
\end{align}
The above proof is non-rigorous and the reader is referred to the Appendix~\ref{sec:singular connections on line bundles associated with generic sections} for a more rigorous proof. Finally, writing $a(\bm{\lambda})=x_1(\bm{\lambda})+ix_2(\bm{\lambda})$, we learn that
\begin{align}
F^{\text{sing}}(\boldsymbol\lambda)=dA^{\text{sing}}(\bm{\lambda})=-2\pi i \delta^2(a(\bm{\lambda}))dx_1(\bm{\lambda})\wedge dx_2(\bm{\lambda}) 
\end{align}
It follows that the curvature of this singular connection, which is defined by means of the additional datum of a section $s$, localizes over its zero locus $Z=\{\bm{\lambda}\in M :s(\bm{\lambda})=0\}$. In particular, it follows that if we take a differential  $\left(m-2\right)$-form $\eta$ in $M$, locally written as
\begin{align}
\eta(\bm{\lambda})=\sum_{i_1,\dots,i_{d-2}}b_{i_1\dots i_{d_2}}(\bm{\lambda})d\lambda_{i_1}\wedge \dots \wedge d\lambda_{i_{d-2}},
\end{align}
for some smooth functions $b_{i_1,\dots, i_{d-2}}(\bm{\lambda})$, we have that
\begin{align}
\int_{M} \frac{iF^{\text{sing}}}{2\pi}\wedge \eta =\int_{Z}\eta,
\end{align}
this statement being consistent with the known fact that the first Chern class of a line bundle is Poincaré dual to the zero locus of a generic section~\cite{griffiths:harris:14,huybrechts:05}.

We point out that there is a natural way to generate global sections of $\mathscr{L}\subset M\times \mathbb{C}^{n+1}$, and hence singular connections on it. We take a generic vector $\ket{\psi_0}\in \mathbb{C}^{n+1}$ and simply consider the assignment $s:\bm{\lambda}\mapsto P(\bm{\lambda})\ket{\psi_0}$, where $P(\bm{\lambda})=\frac{\ket{\psi(\bm{\lambda})}\bra{\psi(\bm{\lambda})}}{\bra{\psi(\bm{\lambda})}\psi(\bm{\lambda})\rangle}$ is the projector onto the quantum state under consideration. By construction, the vector $P(\bm{\lambda})\ket{\psi_0}$ belongs to the fiber at $\bm{\lambda}$ of $\mathscr{L}$. The zero locus of the resulting section $s$ is simply the set of points $\bm{\lambda}$ where $\ket{\psi}$ is orthogonal to the fiber over it, which is equivalent to $\bra{\psi}P(\bm{\lambda})\ket{\psi}=0$. Note that this construction depends on the choice of $\ket{\psi_0}$.

The reader is redirected to  Appendix~\ref{sec:singular connections on line bundles associated with generic sections} for a more detailed mathematical exposition on singular connections on smooth line bundles determined by (generic) sections and the distributional character of the associated curvature.
We have thus introduced the singular connection, locally represented by $A^\mathrm{sing}(\boldsymbol\lambda)$, and the associated singular curvature $F^\mathrm{sing}(\boldsymbol\lambda)$, which are perfectly valid alternatives for the Berry connection and the Berry curvature to study the topology of the line bundle $\mathscr{L}$. In the next two subsections, we demonstrate how the singular connection and curvature behave for two representative examples.

\subsection{Spins in magnetic fields}
\label{subsec: spins in magnetic fields}
Following Michael Berry~\cite{berry:84}, we consider a particle of spin $s$ interacting with a magnetic field via the Hamiltonian
\begin{align}
H(\vec{B})=\vec{B}\cdot \vec{S},
\end{align}
where $\vec{B}=(B_1,B_2,B_3)\in\mathbb{R}^3$ is the magnetic field and $\vec{S}=(S_1,S_2,S_3)$ is a spin $s$ operator. We use natural units to avoid complicating the formulas. The eigenenergies of the above Hamiltonian are 
\begin{align}
|\vec{B}|m, \text{ with } m\in \{-s,\dots, s\}.
\end{align}
We consider $\vec{B}\neq 0$, and denote by $\mathscr{L}_{m}$ the line bundle over parameter space determined by the energy eigenvalue $m|\vec{B}|$.
Here we focus on the simplest case of $s = 1/2$ and the lowest energy eigenbundle $\mathscr{L}_{-\frac{1}{2}}$; general cases for $s \ge 1/2$ are explained in detail in the Appendix~\ref{sec: spins in magnetic fields---general construction}. We may write $\vec{S}=(1/2)\vec{\sigma}$, with $\vec{\sigma}=(\sigma_1,\sigma_2,\sigma_3)$ the Pauli matrices. Additionally, $H(\vec{B})=(1/2)|\vec{B}|\vec{n}\cdot \vec{\sigma}$ with $\vec{n}=\vec{B}/|\vec{B}|$ a unit vector in the Bloch sphere. We can introduce the projector $P(\vec{n})=(1/2)\left(I_2 - \vec{n}\cdot \vec{\sigma}\right)$, where $I_2$ denotes the $2\times 2$ identity matrix, and then the bundle $\mathscr{L}_{-\frac{1}{2}}$ is completely determined by $P(\vec{n})$---the fiber at $\vec{B}$, with $\vec{B}\neq 0$, is the subspace of $\mathbb{C}^2$ in the image of $P(\vec{n})$. Now take a vector $\ket{\psi_0}\in\mathbb{C}^2$, which is determined by a vector $\vec{n}_0=\frac{\bra{\psi_0}\vec{\sigma}\ket{\psi_0}}{\bra{\psi_0}\psi_0\rangle}$ in the Bloch sphere. By the comment in the end of Sec.~\ref{sec: singular connections}, it follows that $P(\vec{n})\ket{\psi_0}$ defines a section $s$ of $\mathscr{L}_{\frac{1}{2}}$ and its zeros are determined by the equation $1-\vec{n}\cdot\vec{n}_0=0$, which follows from $\bra{\psi_0}P(\vec{n})\ket{\psi_0}=0$. It follows that $s$ has one zero corresponding to $\vec{n}=\vec{n}_0$. The curvature of the associated singular connection will be localized there. To relate this to homogeneous polynomials, we note that if we introduce the complex coordinate $z=\frac{n_1+in_2}{1+n_3}=\frac{B_1+iB_2}{|\vec{B}|+B_3}$, valid away from the south pole of the Bloch sphere, then
\begin{align}
\ket{u}=\left[\begin{array}{c}
-\bar{z}\\
1
\end{array}\right],
\end{align}
then $P(\vec{n})\ket{u}=\ket{u}$. Observe then that, writing $\ket{\psi}=(a,b)$, for $a,b\in\mathbb{C}$, we have
\begin{align}
P(\vec{n})\ket{\psi_0}=\left(-az + b \right)\left(\frac{1}{(1+|z|^2)}\ket{u}\right).
\end{align}
We may choose a different eigenvector other than $\ket{u}$ by performing the gauge transformation $\ket{u}\to \lambda \ket{u}$, and then we see that $(-az+b)\to \lambda (-az+b)$. It follows that the section associated with $\ket{\psi_0}$ is completely determined by the homogeneous polynomial $f(z_0,z_1)=-az_0+bz_1$ of degree $2s=1$. The zero of this section is determined by $-a+bz=0\implies z=\frac{a}{b}$ which is precisely the complex coordinate of $\vec{n}_0$. Because the section can be locally described by holomorphic functions, $(-az+b)$, and these are always orientation preserving transformations of the plane, it follows from the results of Sec.~\ref{sec: singular connections} that the associated singular connection satisfies
\begin{align}
F^{\text{sing}}=-2\pi i \delta^2(z-z_1)\frac{i}{2}dz\wedge d\bar{z},
\end{align}
where $\delta^2(z)\frac{ idz\wedge d\bar{z}}{2}$ is such that for a test function $g(z)$ we have $\int_{\mathbb{C}} g(z)\delta^2(z)\frac{ idz\wedge d\bar{z}}{2}=g(0)$. The Berry connection in this case would give
\begin{align}
F=-\frac{i}{2} \vec{n}\cdot\left( d\vec{n}\times d\vec{n}\right),
\end{align}
where $\vec{n}\cdot\left( d\vec{n}\times d\vec{n}\right)$ is the standard area form on the sphere with total area $4\pi$. Both formulas reproduce the result that the first Chern number of $\mathscr{L}_{-\frac{1}{2}}$, when restricted to a sphere enclosing the origin, is equal to $1$.

\begin{figure}[h]
    \centering
    \includegraphics[scale=0.5]{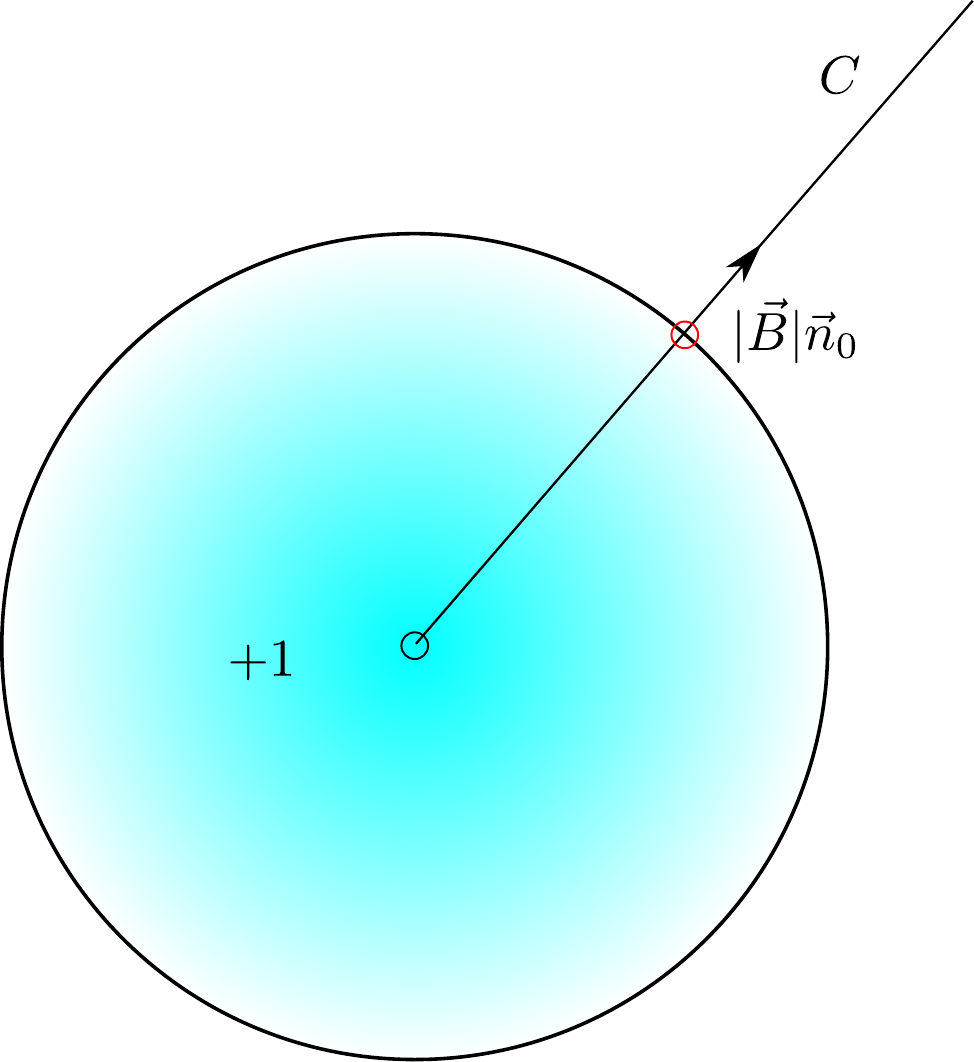}
    \caption{The Dirac string $C$ in the figure is the zero locus of the section which gave rise to the singular connection. Because the first Chern number, the topological charge, is positive $(+1)$, $C$ is oriented so that we may think of it as a string in the parameter space of magnetic fields emanating the location of the monopole $\vec{B}=0$ and going all the way to infinity.}
    \label{fig: Dirac string}
\end{figure}

We point out that the singular connection described here formalizes the concept of a \emph{Dirac string}. The Dirac string should connect two monopoles of opposite ``magnetic charges'' or, when there is only one source, the Dirac string emanates from the monopole and goes to infinity, its direction depending on the sign of the charge of the monopole. Physically, it corresponds to the points where the gauge potential cannot be defined---in our case, the zeros of the section used to define the singular connection. Usually one argues that the position of the Dirac string is gauge dependent and not observable, because one works in the setting of smooth connections and the singularity, which is artificial, arises from trying to extend a gauge which is only valid, locally, in a certain patch. In the present case, since we use a section to define a \emph{singular} connection, as opposed to a \emph{smooth} one, the location of the Dirac string, determined by the zero locus of the section is independent of the gauge. To be more precise, the local representation of the connection as described by $A^\mathrm{sing}(\boldsymbol\lambda)$ does depend on the gauge chosen to represent the section, that is, it changes under the gauge transformation $a(\boldsymbol\lambda) \to g(\boldsymbol\lambda)^{-1} a(\boldsymbol\lambda)$. However, the location of the singularity of $A^\mathrm{sing}(\boldsymbol\lambda)$ does not change, and it is this singularity of $A^\mathrm{sing}(\boldsymbol\lambda)$ which determines the location of the Dirac string.  The Dirac string, in this setting of singular connections, is not only the zero locus of the preferred section which determined the connection, but also the \emph{support}, in parameter space, of the curvature of the connection. To see how singular connections formalize this idea, note that in the parameter space of all nonvanishing magnetic fields $\mathbb{R}^3-\{0\}$, the condition for the zero of the section gives a curve emanating from the origin, where the monopole is located, namely
\begin{align}
C=\{\vec{B}=t\vec{n}_0: t>0\},
\end{align}
and $F^{\text{sing}}$ localizes precisely in $C$, which we refer to as the Dirac string, see Fig.~\ref{fig: Dirac string}. In this case $C$ has to be oriented such that $t$ is in the correct orientation---this is related to the first Chern number, corresponding to the ``magnetic charge'', being positive (otherwise one would choose the opposite orientation for $C$). Now if we take an arbitrary (compact support) one-form $\eta(\vec{B})=\sum_{i=1}^3b_i(\vec{B})dB_i$, for some functions $b_i(\vec{B})$, $i=1,2,3$, we may express it in terms of $r=|\vec{B}|$ and $\vec{n}=\vec{B}/|\vec{B}|$, namely, we will be able to write it as
\begin{align}
\eta(\vec{B})=\sum_{j=1}^3c_j(\vec{B})dn_j +c_r(\vec{B}) dr,
\end{align}
where the above formula is consistent with $\vec{n}\cdot d\vec{n}=0$, so that only three out of the four $c_i(\vec{B})$'s in the formula are really independent. The result of the above computation gives
\begin{align}
\int_{\mathbb{R}^3-\{0\}}\frac{ iF }{2\pi }\wedge \eta =\int_{C}\eta =\int_{0}^{+\infty} c_{r}(-t\vec{n}_0)dt,
\end{align}
so the curvature localizes the integral to be evaluated over $C$.

The more general situation of $\mathscr{L}_{m}$ for $m\in \{-s,\dots, s\}$ with $s \ge 1/2$ is described in Appendix~\ref{sec: spins in magnetic fields---general construction}, corresponds to having  $2|m|$ independent Dirac strings, one for each zero, emanating from the origin. The first Chern number, which is the magnetic charge in the analogy, is obtained by counting the Dirac strings with signs taking orientations into account.

\subsection{Massive Dirac model}
\label{subsec: massive Dirac model}
We next consider the massive Dirac model in two dimensions, which is described by the following two-band Bloch Hamiltonian
\begin{align}
H(\bf{k})=\vec{d}(\bf{k})\cdot\vec{\sigma},
\end{align}
where 
\begin{align}
\vec{d}(\bf{k})=(\sin(k_1),\sin(k_2),M_{D}-\cos(k_1)-\cos(k_2)).
\end{align} 
We now analyze the topology of the lowest energy Bloch bundle, which is a complex line bundle with the base manifold given by the Brillouin zone $\BZ^2$ and the fiber given by the eigenspace of the Hamiltonian $H(\bf{k})$ with the lowest energy. The model has two bands which we will label by $+$ and $-$, with energy dispersions determined by $E_{\pm}=\pm |\vec{d}(\bf{k})|$. As one varies $M_D$, the model undergoes topological phase transitions, with the first Chern number of the lowest band being $0$ for $|M_D|>2$, $1$ for $-2<M_D<0$ and $-1$ for $0<M_D<2$.

Away from the points where $\vec{d}(\bf{k})=0$, considering $M_D \in \mathbb{R}$ also as a parameter,
$\vec{d}$ determines a map from $\BZ^2\times \mathbb{R}$ to $\mathbb{R}^3-\{0\}$. The lowest energy Bloch bundle is determined by the pullback of $\mathscr{L}_{-\frac{1}{2}}$ defined in Sec.~\ref{subsec: spins in magnetic fields}. To be concrete, for fixed $M_D$ with $M_D\neq \{-2,0,2\}$, the vector $\vec{d}(\bf{k})$ determines a map $\vec{d}: \BZ^2\to \mathbb{R}^3-\{0\}$, where $\mathbb{R}^3-\{0\}$ is the parameter space of the model of Sec.~\ref{subsec: spins in magnetic fields}. The lowest energy Bloch bundle has fiber at $\bf{k}$ the eigenspace of $H(\bf{k})$ with eigenenergy $-|\vec{d}(\bf{k})|$, which is precisely the fiber of $\mathscr{L}_{-\frac{1}{2}}\to\mathbb{R}^3-\{0\}$ at the image of $\bf{k}$ by the map $\vec{d}$, i.e., at $\vec{B}=\vec{d}(\bf{k})$. The section of $\mathscr{L}_{-\frac{1}{2}}$ determined by $\ket{\psi_0}=(1,0)$ through projection has a zero at the north pole of the Bloch sphere. To determine the curvature, we need to look at the points of $(\bf{k},M_D)$ of parameter space that are mapped to the north pole of the Bloch sphere, i.e., where $\vec{n}(\bf{k})=\frac{\vec{d}(\bf{k})}{|\vec{d}(\bf{k})|}=(0,0,1)$. These are curves, conveniently parameterized by $M_D$, given by
\begin{align}
C_1 &=\{(\pi,\pi,M_D): M_D>-2\}, \nonumber \\
C_2 &=\{(\pi, 0, M_D): M_D>0\}, \nonumber \\
C_3 &=\{(0,\pi, M_D): M_D>0\}, \nonumber\\
C_4 &=\{(0,0,M_D): M_D>2\}.
\end{align}
The orientations of each of these curves is prescribed according to whether $\vec{n}$, restricted to any point in the curve, is orientation preserving or reversing. It can be effectively calculated from the formula $\sigma_{\bf{k}}=\text{sign}\left[\vec{d}\cdot \left(\frac{\partial \vec{d}}{\partial k_1}\times \frac{\partial \vec{d}}{\partial k_2} \right) \right]$. Accordingly, $M_D$ is a positively oriented coordinate for $C_1$ and $C_4$, and negative for $C_2$ and $C_3$. The curvature assumes the expression
\begin{widetext}
\begin{align}
&\frac{i F^{\text{sing}}}{2\pi}\nonumber\\
&=\left[\theta(M_D+2)\delta^2(\bf{k}-(\pi,\pi))dk_1\wedge dk_2 -\theta(M_D)\left(\delta^2(\bf{k}-(0,\pi))+\delta^2(\bf{k}-(\pi,0))\right)dk_1\wedge dk_2+\theta(M_D-2)\delta^2(\bf{k})dk_1\wedge dk_2 \right],
\end{align}
\end{widetext}
where $\theta(\cdot)$ denotes the Heaviside step function. Observe also that for fixed $M_D$, $M_D\neq \{-2,0,2\}$, this formula retrieves the usual expression of the first Chern number in terms of the degree or winding number of the map defined by $\vec{n}$:
\begin{align}
\int_{\BZ^2} \frac{i F^{\text{sing}}}{2\pi}=\sum_{\{\bf{k}: \vec{n}(\bf{k})=(0,0,1)\}}\sigma_{\bf{k}},
\end{align}
since the right-hand side is simply counting the points in the pre-image of $(0,0,1)$ according to whether $\vec{n}:\BZ^2\to S^2$ is orientation preserving or reversing at those points. 

\begin{figure}[h!]
    \centering
    \includegraphics[scale=0.5]{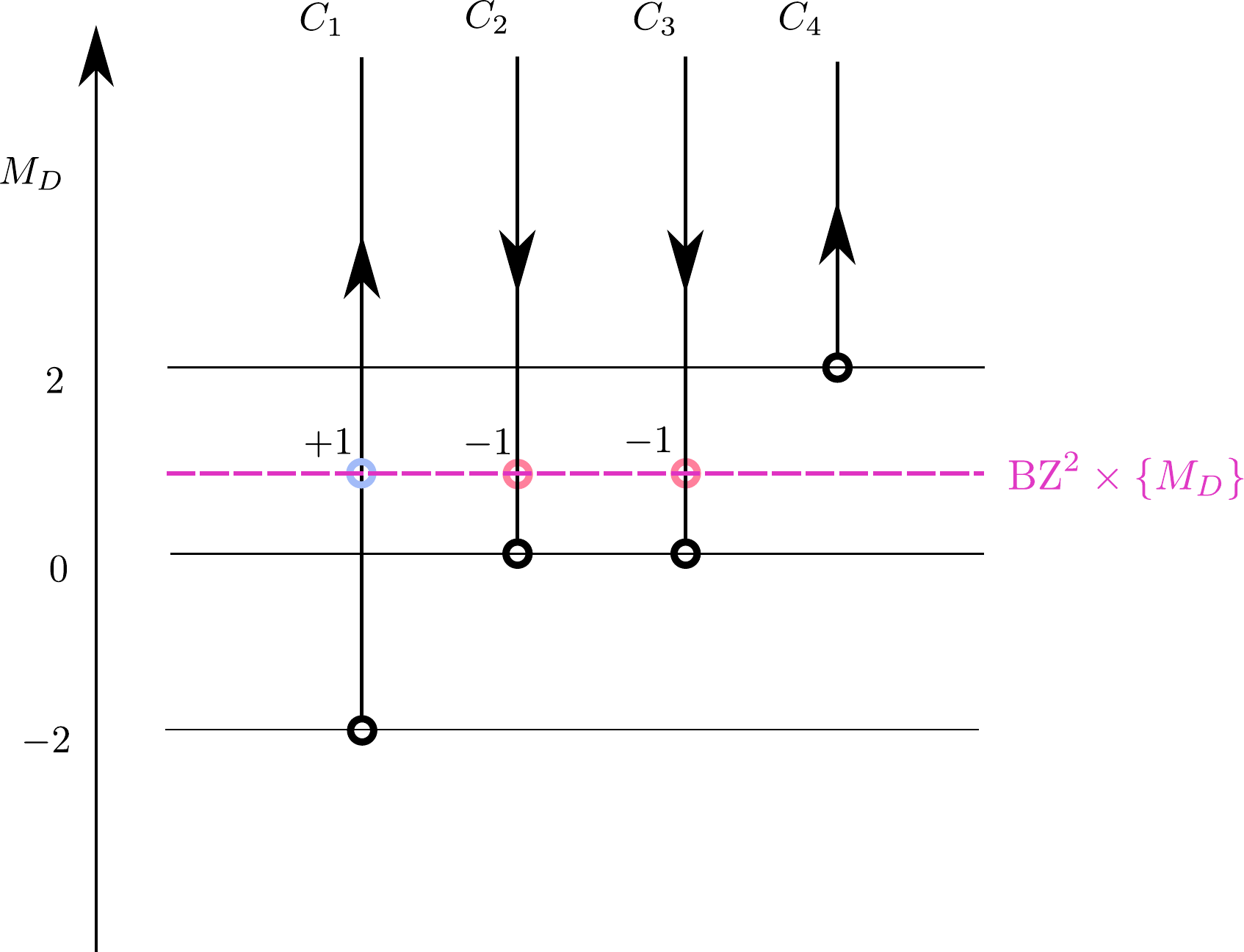}
    \caption{As we change $M_D$, the submanifold $\BZ^2\times \{M_D\}$ intersects the curves $C_1,C_2,C_3,C_4$, which one may interpret as Dirac strings, in the parameter space ($\BZ^2\times \mathbb{R}$ minus the points of phase transition) differently. As we reach the critical values $M_D\in \{-2,0,2\}$ the intersection number jumps by an integer, because $\BZ^2\times \{M_D\}$ intersects more Dirac strings. The signs $\pm 1$, and also the respective colors blue and magenta, indicate whether the intersection of $\BZ^2\times \{M_D\}$ and the $C_i$'s, when nonempty, occurs in the positive or negative orientation of the ambient manifold. In the figure, the first Chern number equals $+1-1-1=-1$, which is consistent with the region of the phase diagram.}
    \label{fig: massive Dirac model topological phase transitions}
\end{figure}

In terms of the above formula for the curvature, the topological phase transitions can be understood very simply. For fixed $M_D$ the submanifold $\BZ^2\times \{M_D\}$ of parameter space will have an intersection number, which counts the number of intersection points taking into account the orientation (which here is expressed by weighting them with the signs $\sigma_{\bf{k}}$ at the intersection points), with $\bigcup_{i}C_i$. This intersection number is precisely the first Chern number. As we change $M_D$, the Chern number changes simply because the number of curves in $\bigcup_{i}C_i$ that $\{M_D\}\times \BZ^2$ intersects changes. In light of the discussion in Sec.~\ref{subsec: spins in magnetic fields}, we may interpret the $C_i$'s as Dirac strings, and the first Chern number as the intersection number of the Brillouin zone, embedded in the extended parameter space, with the Dirac strings. In Fig.~\ref{fig: massive Dirac model topological phase transitions}, we illustrate this idea. Additionally, in the discussion of Sec.~\ref{subsec: spins in magnetic fields}, we could have also calculated the first Chern number of $\mathscr{L}_{\frac{1}{2}}$ restricted to an embedded sphere of the form $\vec{n}\mapsto \vec{B}=c\vec{n}$, where $c$ is a real number. The calculated Chern number is equal to $-1$, independently of $c$ as found before, which is now also clear from Fig.~\ref{fig: Dirac string}, since the sphere would always intersect the Dirac string exactly once.

\section{Singular connection for a pair of quantum states}
\label{sec: transition dipoles}
We now consider a pair of quantum states and transitions between the pair. As we explain below, the setup can also be described in terms of a complex line bundle over a parameter space, and the singular connection can be introduced exactly as in the previous section.

We consider two families of quantum states, $|\psi_1 (\boldsymbol\lambda)\rangle$ and $|\psi_2 (\boldsymbol\lambda)\rangle$. We have corresponding line bundles $\mathscr{L}_1 \to M$ and $\mathscr{L}_2 \to M$. From these line bundles, we can define the {\it homomorphism bundle} by $\mathscr{L}_{12} := \mathscr{L}_1 \otimes \mathscr{L}_2^* = \mathrm{Hom}(\mathscr{L}_2, \mathscr{L}_1)$, which is nothing but the line bundle whose fiber at $\boldsymbol\lambda \in M$ is given by the one dimensional complex vector space spanned by $|\psi_1 (\boldsymbol\lambda)\rangle\langle \psi_2 (\boldsymbol\lambda)|$. Here, $\mathscr{L}_{2}^*$ denotes the dual bundle of $\mathscr{L}_{2}$ whose fiber at $\boldsymbol\lambda \in M$ is the dual of the fiber of $\mathscr{L}_2$ at that point.
For the complex line bundle $\mathscr{L}_{12} \to M$, we can define the singular connection as before once a global section is given. As we discuss, there are cases where natural global sections exist for the homomorphism bundle $\mathscr{L}_{12}$, and thus we can attribute physical meaning to the singular connection and curvature. Most relevant is the situation where $|\psi_1 (\boldsymbol\lambda)\rangle$ and $|\psi_2 (\boldsymbol\lambda)\rangle$ are both eigenstates of a Hamiltonian $H(\boldsymbol\lambda)$. By adding external perturbation, we can induce transition from $|\psi_1 (\boldsymbol\lambda)\rangle$ to $|\psi_2 (\boldsymbol\lambda)\rangle$. As we will see, the transition matrix element gives a preferred global section.

To make the discussion concrete, we consider the case where the parameter space is the Brillouin zone, $\boldsymbol\lambda = \mathbf{k}$, and the pair of quantum states are the Bloch wave functions for nondegenerate bands $m$ and $n$, locally described by $\ket{u_{m,\bf{k}}}$ and $\ket{u_{n,\bf{k}}}$. We assume that these Bloch wave functions are normalized.
The Bloch wave functions for bands $m$ and $n$ define the line bundles $\mathscr{L}_{m}\to\BZ^2$ and $\mathscr{L}_n\to \BZ^2$, respectively, and the associated homomorphism bundle  $\mathscr{L}_{mn} =\text{Hom}(\mathscr{L}_{n},\mathscr{L}_m)$ whose fibers are spanned by $\ket{u_{m,\bf{k}}}\bra{u_{n,\bf{k}}}$.
The transition dipole matrix element for optical transitions between the two bands for linearly polarized light along $x_j$ direction is given by $r_{mn}^j = i\langle u_{m,\mathbf{k}}|\frac{\partial}{\partial k_j}|u_{n,\mathbf{k}}\rangle$~\cite{aversa:1995}. This transition dipole matrix element defines a global section of $\mathscr{L}_{mn}$ given by
\begin{align}
    s^{j}(\bf{k})&=r^{j}_{mn}(\bf{k})\ket{u_{m,\bf{k}}}\bra{u_{n,\bf{k}}}
\nonumber \\
 &=i\bra{ u_{m,\bf{k}}}\frac{\partial}{\partial k_j}\ket{u_{n,\bf{k}}}\ket{u_{m,\bf{k}}}\bra{u_{n,\bf{k}}}.
\end{align}
One can see that $s^{j}(\bf{k})$ defines a valid global section by explicitly checking the invariance of $s^{j}(\bf{k})$ upon gauge transformations $\ket{u_{m,\bf{k}}}\to g_m(\bf{k}) \ket{u_{m,\bf{k}}}$, $\ket{u_{n,\bf{k}}}\to g_n(\bf{k}) \ket{u_{n,\bf{k}}}$ for $\text{U}(1)$-valued gauge transformations $g_m$ and $g_n$.

The meaning of $s^j$ being a section of the homomorphism bundle $\mathscr{L}_{mn}$ is very simple and intuitive: the transition dipole describes transition amplitudes between bands $n$ and $m$---hence, it naturally defines, for each $\bf{k}\in\BZ^2$, a linear map from the fiber at $\bf{k}$ of $\mathscr{L}_n$ to the fiber at $\bf{k}$ of $\mathscr{L}_m$. The line bundle $\mathscr{L}_{mn}$ carries a connection induced from the Berry connections on $\mathscr{L}_m$ and $\mathscr{L}_n$. On the local gauge given by $\ket{u_{m,\bf{k}}}\bra{u_{n,\bf{k}}}$ it is described the connection $1$-form
\begin{align}
A_{mn}=A_m-A_n,    
\end{align}
with
\begin{align}
A_m=\bra{u_{m,\bf{k}}}d\ket{u_{m,\bf{k}}}, \text{ and } A_n=\bra{u_{n,\bf{k}}}d\ket{u_{n,\bf{k}}},
\end{align}
being the local connection forms for the Berry connections on $\mathscr{L}_m$ and $\mathscr{L}_n$, respectively.

Now the fact that we have \emph{a preferred nontrivial section} of $\mathscr{L}_{mn}$ allows us to define a singular connection on it by declaring
\begin{align}
\nabla s^j\equiv 0.
\end{align}

Since $\ket{u_{m,\bf{k}}}\bra{u_{n,\bf{k}}}=s^j \left(r^{j}_{mn}\right)^{-1}$, using the Leibniz rule, we can show that the connection coefficient on the gauge $\ket{u_{m,\bf{k}}}\bra{u_{n,\bf{k}}}$ is
\begin{align}
A^{\text{sing},j}_{mn}=-d\ln r^{j}_{mn},
\end{align}
The difference between two connections always defines a $1$-form, which in this case, because one of them is singular, defines a $1$-current in the sense defined by de Rham~\cite{deRham:12} (see Appendix~\ref{sec:singular connections on line bundles associated with generic sections} for more details):
\begin{align}
\delta^{j}_{mn}:=A_{mn}-A^{\text{sing},j}_{mn}=A_m-A_n +d\ln r^{j}_{mn}. 
\end{align}
This 1-current, up to an overall factor of $i$, is nothing but the {\it shift vector}, measuring the shift of the electron position during the optical excitation from band $n$ to band $m$~\cite{ahn:guo:nagaosa:vishwanath:22}. The fact that the shift vector is constructed from the difference of two connections gives the geometrical interpretation of the gauge invariance of the shift vector. We note that there is also a convention to call only the imaginary part of $\delta^j_{mn}$ as the shift vector~\cite{sipe:shkrebtii:2000, morimoto:nagaosa:16, nagaosa:morimoto:17}, but this difference does not affect the results we present.

In Ref.~\cite{ahn:guo:nagaosa:vishwanath:22}, a topological invariant for the optical transition between two bands is introduced, which is the difference of the Chern numbers of the two bands. This topological invariant can be regarded as the Chern number of the homomorphism bundle $\mathscr{L}_{mn}$, which is obtained by integrating the curvature $ iF_{mn}/(2\pi)=idA_{mn}/(2\pi)$ over the Brillouin zone. Since $A_{mn}$, obtained from the Berry connections of each band, and the singular connection $A_{mn}^{\mathrm{sing},j}$ are both valid connections in the homomorphism bundle, we can alternatively compute the same topological invariant through the singular connection $A_{mn}^{\mathrm{sing},j}$ by integrating its associated singular curvature. The singular curvature takes an infinite value, in the sense of the Dirac delta function, where the global section used for its construction becomes zero, which is nothing but the zeros of the transition dipole matrix elements $r^{j}_{mn}$. Concretely, by decomposing $r^j_{mn}(\bf{k})$ into real and imaginary parts, $r^j_{mn}(\bf{k})=x_1(\bf{k})+ix_2(\bf{k})$, we can write
\begin{align}
F^{\text{sing},j}=-2\pi i\delta^{2}(r^j_{mn}(\bf{k}))dx_1(\bf{k})\wedge dx_2(\bf{k}).    
\end{align}
In the case of two dimensions, where the zero locus $Z$ of $r^j_{mn}$ is (generically) just a finite collection of points, we can simplify the above expression to find the simple formula
\begin{align}
F^{\text{sing},j}=\sum_{\bf{k}_*\in Z}\sigma_{\bf{k}_*}\delta^2(\bf{k}-\bf{k}_*)dk_1\wedge dk_2,    
\end{align}
where $\sigma_{\bf{k}_*}\in\{\pm 1\}$ takes into account whether $r^j_{mn}$ determines an orientation preserving or reversing map in a small neighbourhood of $\bf{k}_*$, $\bf{k}_*\in Z$. The integral of the singular curvature algebraically counts the number of points in the Brillouin zone where transitions between band $m$ and $n$ are not allowed via $r^{j}_{mn}$, i.e., via the absorption of a photon. Note that this number is independent of $j$, so for the computation of the topological invariant it does not matter which direction we choose to define the singular connection.

In the next section, we illustrate the results of this section by performing computations on two explicit models.
\subsection{Transitions in the Bloch sphere}
\label{subsec: transition dipoles in the Bloch sphere}
Instead of starting with the Brillouin zone as the parameter space, we first assume a toy model for demonstration where we take the base space to be a two-dimensional sphere, in which case analytical calculation of the singular connection is possible. We consider the Hamiltonian
\begin{align}
H(\theta, \varphi)= - \begin{pmatrix} \cos \theta & \sin \theta e^{-i\varphi} \\ \sin \theta e^{i\varphi} & -\cos \theta \end{pmatrix} = -\vec{n}\cdot \vec{\sigma},
\end{align}
where $\vec{n} = (\sin \theta \cos \varphi, \sin \theta \sin \varphi, \cos \theta) \in S^2 \subset \mathbb{R}^3$ is a vector at a point on the Bloch sphere.
Such a Hamiltonian can be realized in various engineered quantum systems and the quantum geometry of such a Hamiltonian has been fully characterized, for example, in diamond NV centers~\cite{Yu:2020}, superconducting qubits~\cite{Tan:2019}, and multiterminal Josephson junctions~\cite{Klees:2020}. By periodically perturbing the Hamiltonian, transition between the two eigenstates, $|u_-\rangle$ and $|u_+\rangle$ can be induced. If one, for example, varies $\varphi$ sinusoidally, the transition matrix element is proportional to $\langle u_+|\frac{\partial}{\partial \varphi}| u_-\rangle$~\cite{Ozawa:2018}, which we can use as a natural global section as we see below.

This Hamiltonian is equivalent to that considered in Sec.~\ref{subsec: spins in magnetic fields}, where we take $s=1/2$ and $|\vec{B}|=1$. Accordingly, we have two line bundles $\mathscr{L}_{\pm}$, resulting from the restrictions of $\mathscr{L}_{\mp\frac{1}{2}}$ to $\vec{B}=1$, associated to eigenstates of energy $\pm 1$ and whose fibers at $\vec{n}$ are determined by the image of the projectors $P_{\pm}(\vec{n})=(1/2)(I_2 \pm H(\vec{n}))$, where $I_2$ is the $2\times 2$ identity matrix. The Berry connections associated to $\mathscr{L}_{\pm}$ give the following Berry curvatures
\begin{align}
F_{\pm}=\mp \frac{i}{2}\vec{n}\cdot \left(d\vec{n}\times d\vec{n}\right).
\end{align}
It follows that the first Chern numbers of $\mathscr{L}_{\pm}$ are, respectively, $\pm 1$. As a consequence, the first Chern number of $\mathscr{L}_{+}\otimes \mathscr{L}_{-}^*$ is simply $1-(-1)=2$.
We can take local Bloch wave functions for $\mathscr{L}_{\pm}$, defined on $S^2-\{(0,0,-1)\}$, to be
\begin{align}
\ket{u_{-}} &=\frac{1}{(1+|z|^2)^{1/2}}\left[\begin{array}{c}
1\\
z
\end{array}\right]\nonumber \\
\ket{u_{+}} &=\frac{1}{(1+|z|^2)^{1/2}}\left[\begin{array}{c}
-\bar{z}\\
1
\end{array}\right],
\label{eq: north gauge}
\end{align}
where $z=(n^1+in^2)/(1+n^3) = \tan \left( \frac{\theta}{2} \right) e^{i\varphi}$ is the stereographic projection coordinate, defined with respect to the south pole. We can use the transition matrix element $\langle u_+|\frac{\partial}{\partial \varphi}| u_-\rangle$ upon sinusoidal modulation of $\varphi$ as a section:
\begin{align}
s^{\varphi}(\vec{n}) &= r^{\varphi}_{+-}\ket{u_{+}}\bra{u_{-}} =i\bra{u_{+}}\frac{\partial}{\partial \varphi}\ket{u_{-}}\ket{u_{+}}\bra{u_{-}} \nonumber \\
&=-\frac{ z}{1+|z|^2}\ket{u_{+}}\bra{u_{-}}.
\label{eq: sphi north gauge}
\end{align}
This expression reveals that there is a zero at $z=0$, i.e., at the north pole, and since $z$ winds in the positive orientation of the complex plane, we assign $+1$ to this zero (see Appendix~\ref{sec:singular connections on line bundles associated with generic sections}). To probe what happens near the south pole, we introduce the complex coordinate $w=1/z=(n^1-in^2)/(1-n^3)$ which is the stereographic projection coordinate defined respect to the north pole and is well-defined away from it. We can then write new local gauges
\begin{align}
\ket{u'_{-}} &=\frac{1}{(1+|w|^2)^{1/2}}\left[\begin{array}{c}
w\\
1
\end{array}\right]\nonumber \\
\ket{u'_{+}} &=\frac{1}{(1+|w|^2)^{1/2}}\left[\begin{array}{c}
1\\
-\bar{w}
\end{array}\right],
\label{eq: south gauge}
\end{align}
valid now over $S^2-\{(0,0,1)\}$. We can then obtain an expression for the transition dipole in this new trivialization,
\begin{align}
s^{\varphi}(\vec{n}) &= r'^{\varphi}_{+-}\ket{u'_{+}}\bra{u'_{-}} =i\bra{u'_{+}}\frac{\partial}{\partial \varphi}\ket{u'_{-}}\ket{u'_{+}}\bra{u'_{-}} \nonumber \\
&=\frac{w}{1+|w|^2}\ket{u'_{+}}\bra{u'_{-}}.
\label{eq: sphi south gauge}
\end{align}
Note that on the overlap of the two neighbourhoods, i.e., $S^2-\{(0,0,-1),(0,0,1)\}$ we have
\begin{align}
\ket{u_{-}}=\frac{z}{|z|}\ket{u'_{-}}, \ket{u_{+}}=-\frac{\bar{z}}{|z|}\ket{u'_{+}},
\end{align}
and also 
\begin{align}
-\frac{z}{1+|z|^2}=-\frac{\bar{w}}{1+|w|^2}.
\end{align}
It follows that
\begin{align}
-\frac{z}{1+|z|^2}\ket{u_{+}}\bra{u_{-}} &=-\frac{\bar{w}}{1+|w|^2}\left(-\frac{\bar{z}}{|z|}\right)\frac{\bar{z}}{|z|}\ket{u'_{+}}\bra{u'_{-}} \nonumber \\
&=\frac{w}{1+|w|^2}\ket{u'_{+}}\bra{u'_{-}},
\end{align}
confirming that $s^{\varphi}(\vec{n})$ is well-defined, i.e., gauge invariant.

With this expression we find that there is a zero at $w=0$, i.e. at the south pole, and since $w$ winds in the positive orientation of the complex plane, we also assign $+1$ to this zero. Since we did this for an open cover of $S^2$, these are the only zeros of the transition dipole section $s^{\varphi}$. This information, together with the results of Sec.~\ref{sec: singular connections}, allows us to write $F^{\text{sing}}$ explicitly as
\begin{align}
F^{\text{sing}}=-2\pi i \delta^2(z) \frac{i}{2}dz\wedge d\bar{z} -2\pi i \delta^2(w) \frac{i}{2}dw\wedge d\bar{w},
\end{align}
i.e., it is supported at the north and south poles of the spheres, determined, respectively, by the equations $z=0$ and $w=0$. We can also conclude that the associated topological invariant, i.e., the first Chern number of $\mathscr{L}_{+}\otimes \mathscr{L}_{-}^{*}$, is given by the sum of the zeros of $s^{\varphi}$ with orientations taken into account, i.e.: $+1+1=2$. This is consistent with the previous reasoning and also with a computation using the connection induced from the Berry connections of $\mathscr{L}_{+}$ and $\mathscr{L}_{-}$, which has curvature
\begin{align}
F_{+-}=F_{+} -F_{-}=- i \vec{n}\cdot \left( d\vec{n}\times d\vec{n}\right),
\end{align}
yielding the same result $\int_{S^2} i F_{+-}/(2\pi)=2$.

\subsection{Transition dipoles in the massive Dirac model}
\label{subsec: Transition dipoles in the massive Dirac model}
Finally, we consider the massive Dirac model in two dimensions, as in Sec.~\ref{subsec: massive Dirac model}. The transition dipole in the $j$th direction
\begin{align}
s^j(\bf{k})&=r^{j}_{+-}(\bf{k})\ket{u_{+,\bf{k}}}\bra{u_{-,\bf{k}}} \nonumber\\
&=i\bra{u_{+,\bf{k}}}\frac{\partial}{\partial k_i}\ket{u_{-,\bf{k}}}\ket{u_{+,\bf{k}}}\bra{u_{-,\bf{k}}}
\end{align}
determines a section of $\mathscr{L}_{+}\otimes \mathscr{L}_{-}$ since it defines transitions from the lowest to the highest energy band. This bundle has a natural Hermitian metric for which
\begin{align}
|| s^j(\bf{k})||^2 &=\tr\left( s^j(\bf{k})^\dagger s^j(\bf{k})\right)=r^{j}_{-+}(\bf{k})r^{j}_{+-}(\bf{k}) \nonumber\\
&=g_{jj}(\bf{k})=\frac{1}{4}\frac{\partial \vec{n}(\mathbf{k})}{\partial k_j}\cdot\frac{\partial \vec{n}(\mathbf{k})}{\partial k_j},
\end{align}
where we noted that, in two-band models, $r^{j}_{-+}(\bf{k})r^{j}_{+-}(\bf{k})$ is precisely the $jj$ component of the quantum metric of the lowest band, defined by~\cite{Provost:1980}
\begin{align}
    g_{jj}(\mathbf{k}) &= \left\langle \frac{\partial u_-}{\partial k_j }\right|\left. \frac{\partial u_-}{\partial k_j }\right\rangle - \left\langle \left.\frac{\partial u_-}{\partial k_j }\right| u_-\right\rangle \left\langle u_- \left| \frac{\partial u_-}{\partial k_j }\right.\right\rangle
    \notag \\
    &=
    \left\langle \left.\frac{\partial u_-}{\partial k_j }\right| u_+\right\rangle \left\langle u_+ \left| \frac{\partial u_-}{\partial k_j }\right.\right\rangle.
\end{align}
In the previous equation $\vec{n}(\bf{k})=\vec{d}(\bf{k})/|\vec{d}(\bf{k})|$ is the vector on the Bloch sphere whose winding over the Brillouin zone determines the topological features of the line bundles $\mathscr{L}_{+}$ and $\mathscr{L}_-$. It follows that looking for zeros of $s^j(\bf{k})$ is equivalent to looking for zeros of $g_{jj}(\bf{k})$ which, in turn, correspond to the points where $\frac{\partial \vec{n}}{\partial k_j}=0$.
Below, we numerically plot the quantum metric $g_{jj}(\bf{k})$ to find the zeros of $s^j(\bf{k})$. We note that the transition dipole matrix element itself, $s^j(\bf{k})$, is a gauge dependent quantity, and, in general, cannot be plotted globally. On the other hand, the quantum metric $g_{jj}(\bf{k})$ is itself a gauge invariant quantity and it is possible to plot.

\begin{figure*}[!ht]
\centering
\includegraphics[width = 2.0\columnwidth]{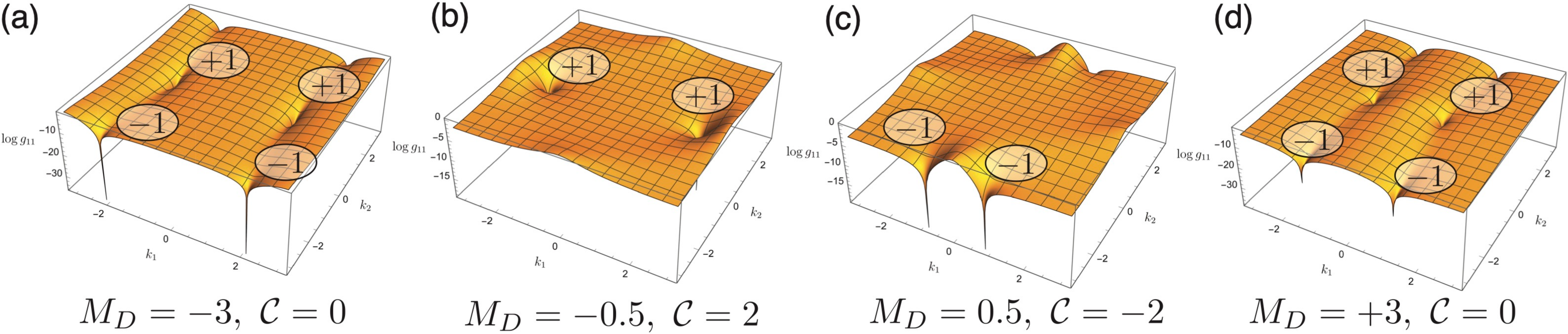}
\caption{Plots of $\ln g_{11}$ as a function of momenta in $\mathrm{BZ}^2$ for different values of $M_D$. Values of $M_D$ and the associated Chern numbers are shown below each panel. Points where $g_{11} = 0$ appear as diverging points in the plot of $\ln g_{11}$; these points corresponds to the zeros of the section $s^1$. Signs $\sigma_\mathbf{k}$ attached to each zeros of $s^1$ according to the orientation of the map $\mathbf{k} \to \vec{n}(\mathbf{k})$ are shown beside each point with $g_{11} = 0$ in the figure. Note that the sum of $\sigma_\mathbf{k}$ of all the points with $g_{11} = 0$ agrees with the Chern number.}
\label{fig:mdm}
\end{figure*}

The Chern number $\mathcal{C}$ of the homomorphism bundle $\mathscr{L}_{+}\otimes \mathscr{L}_{-}^*$ places a constraint on the number of zeros of the transition dipole section since it is the weighted sum of zeros according to the orientation, denoted $\sum_{\bf{k}\in Z} \sigma_\bf{k}$, with $Z=\{\bf{k}\in\BZ^2: s^j(\bf{k})=0\}$ and $\sigma_{\bf{k}}\in \{\pm 1\}$, and this sum satisfies 
\begin{align}
|\mathcal{C}| = \left|\sum_{\bf{k}\in Z}\sigma_{\bf{k}}\right|\leq \sum_{\bf{k}\in Z} 1=\# Z=\text{ number of zeros}.
\end{align}
This means that when the first Chern number of $\mathscr{L}_{+}\otimes \mathscr{L}_{-}^*$ is $\mathcal{C}$, we expect at least $|\mathcal{C}|$ zeros, which we confirm in Fig.~\ref{fig:mdm} by plotting $\ln g_{11} =\ln ||s^1||^2$ as a function of the momenta in the Brillouin zone $\BZ^2$. The zeros of $s^1$ can be identified in the plots as the large (infinite) negative values of $\ln g_{11}$.

For each zero of $s^1$, we can assign the sign $\sigma_{\bf{k}}$, which can be computed by explicitly evaluating the signs
\begin{align}
\sigma_{\bf{k}}=\text{sgn}\left[\frac{i}{2}\left(\frac{\partial r^1_{+-}}{\partial k_1}\frac{\partial \overline{r^1_{+-}}}{\partial k_2}-\frac{\partial r^1_{+-}}{\partial k_2}\frac{\partial \overline{r^1_{+-}}}{\partial k_1}\right)\right],   
\end{align}
for each zero. The results are written in Fig.~\ref{fig:mdm} beside each zero of $g_{11}$. As expected, the sum of $\sigma_{\bf{k}}$ for all the zeros in the Brillouin zone equals to the Chern number $\mathcal{C}$ of the homomorphism bundle. We note that, even for $M_D = \pm 3$ where the Chern number is zero, there are zeros of $s^1$, but the signs cancel each other and their sum is zero as expected. For the cases with $\mathcal{C} = 0$, although the particular section we chose from the transition dipole matrix elements have zeros, there exists a section without zeros since a smooth line bundle with zero first Chern class is trivializable. This is to be contrasted with the cases where $\mathcal{C} \neq 0$, which do not allow for a section without zeros.

\section{Conclusion}
\label{sec: conclusion}
We have introduced singular connections, as determined by the choice of a global section, as an alternative to the Berry connection for families of quantum states. The singular connections have curvature which is localized in the zero locus of the section. We have illustrated the use of these connections for a spin coupled to a magnetic field, comparing with the known results for the Berry curvature and retrieved the first Chern number of the associated bundles, which is independent of the chosen connection. We have also shown the relation of singular connections to the concept of Dirac strings and how one can use it to describe topological phase transitions in two dimensions.

We have described how singular connections appear naturally in the context of transition dipoles and in the definition of the shift vector. The singular connections and the one induced from the Berry connections of each band provide a new natural justification for the gauge invariance of the shift vector. Using the theory of singular connections, we have shown that the topological invariant in two dimensions associated with optical transitions between the two bands can be computed by algebraically counting the points in the zero locus of a transition dipole matrix element of the two bands involved. We have illustrated our results by considering a Bloch sphere toy model and the massive Dirac model. In the process, we have shown how the topological invariant associated with optical transitions provides a natural topological lower bound on the number of momenta in the Brillouin zone for which an electron cannot be excited from one Bloch band to the other by absorbing a photon.

In general, the present work illustrates the idea that, whenever we have a physical situation where there exists a preferred section of a complex line bundle over some parameter space, it is justified to look at the corresponding singular connection. The case of nonlinear optical responses provides an example of such physical situation, where the shift vector is the physical manifestation of the singular connection. The \emph{geometrical amplitude factor or quantum geometric potential}, whose expression is identical to that of the shift vector, except that the parameter space is not necessarily the Brillouin zone, appears in the context of nonadiabatic geometric effects in quantum tunneling~\cite{takayoshi:wu:oka:21}. Therefore, singular connections seem to be present in the general context of transitions between quantum states, where transition dipole-like sections are natural. Transitions between quantum states are usually associated with non-adiabatic processes. This suggests that the physical nature of these singular connections is quite different from the Berry connection, which appears in the adiabatic setting. Finally, one can imagine that the present construction may be generalized to the setting of transitions among more than two bands, which, mathematically, amounts to considering a rank $r>1$ complex vector bundle on a parameter space describing an $r$-fold degenerate family of quantum states. In that case, a natural generalization would be to take $r$ generically linearly independent sections to fix a singular connection. One then expects the curvature to localize in the locus where the $r$ sections fail to be linearly independent. We believe that the singular connections offer natural geometrical language to describe transitions between different quantum states, including systems with degenerate bands, whose study is left for future work.

\begin{acknowledgments}
B.~M. is very grateful to João P. Nunes for going through the proof in Appendix~\ref{sec:singular connections on line bundles associated with generic sections}. B.~M. thanks João E. H. Braz for fruitful discussions. This work is supported by JSPS KAKENHI Grant No. JP20H01845, JST PRESTO Grant No. JPMJPR19L2, and JST CREST Grant No.JPMJCR19T1.
\end{acknowledgments}

\appendix
\section{Singular connections on line bundles associated with generic sections}
\label{sec:singular connections on line bundles associated with generic sections}
We give here a more rigorous exposition of the singular connection for mathematically oriented readers.

If we have a line bundle $\mathscr{L}\to M$ over an oriented smooth manifold $M$ of dimension $m$, together with a smooth section $s$ (other than the trivial zero section), we can define a connection $\nabla$ by declaring
\begin{align}
\nabla s\equiv 0.
\end{align}
This gives rise to a singular connection, as we shall see. Suppose we cover $M$ by open sets $\{U_{\alpha}\}$ where we have trivializing sections $\{e_{\alpha}\}$. Over the $U_{\alpha}$'s we will be able to write
\begin{align}
s=e_{\alpha}f_{\alpha},
\end{align}
for some functions $f_{\alpha}$'s defined over the $U_{\alpha}$'s. The local connection coefficients, which are $1$-forms $\omega_{\alpha}$ defined on the $U_\alpha$'s, will be given by, applying the definition and Leibniz rule,
\begin{align}
\nabla e_{\alpha}=e_{\alpha}\omega_{\alpha} \text{ with } \omega_{\alpha}=-d\ln f_{\alpha}.
\end{align}
The functions $f_{\alpha}: U_{\alpha}\to \mathbb{C}$ define local maps to $\mathbb{C}\cong \mathbb{R}^2$. Note that $-d\ln f_{\alpha}$ is simply the pullback by $f_{\alpha}$ of the meromorphic $1$-form $-d\ln z=-\frac{dz}{z}$. Let us introduce real coordinates in $\mathbb{C}$ such that $z=x^1+ix^2$. Over $\mathbb{C}-\{0\}$ the differential form $d\ln z=\frac{dz}{z}$ is smooth and it satisfies $d^2\ln z=0$. Note, however, that the logarithm of $z$ is multivalued, and we have
\begin{align}
\int_{\gamma}\frac{dz}{z}= 2\pi i\neq 0 ,   
\end{align}
for any contour $\gamma$ encircling the origin of $\mathbb{C}$ once and with the appropriate anti-clockwise orientation. The above result just reflects the fact that the notation $d\ln z$ is misleading, since $\ln z$ is not a smooth function on $\mathbb{C}-\{0\}$. If we tried to extend $dd\ln z$ to the origin we would not be able in the standard sense of a smooth differential $2$-form. Nevertheless, it is possible to interpret it as a $2$-current, in the sense introduced by de Rham~\cite{deRham:12}. Namely, we will take a compactly supported smooth function $f$ in the plane, and define the evaluation of $dd\ln z$ on $f$ to be the integral
\begin{align}
dd\ln z [f]=\int_{\mathbb{C}}d\ln z\wedge df.
\end{align}
This is the differential form analog of the differentiation formula $T'[f]=-T[f']$ which holds for distributions $T$ over $\mathbb{R}$, i.e. $0$-currents, and compactly supported smooth functions $f$ over $\mathbb{R}$. The motivation for the formula is that for smooth $2$-forms of the form $d\beta$, where $\beta$ is a $1$-form, we can use $d(\beta f)=d\beta f -\beta \wedge df$ and integration by parts to show
\begin{align*}
d\beta [f]=\int_{\mathbb{C}} \beta\wedge df =-\int_{\mathbb{C}}d(\beta f) +\int_{\mathbb{C}} d\beta f=\int_{\mathbb{C}} d\beta f,
\end{align*}
since $f$ has compact support. The right-hand side is the natural linear functional on the space of compactly supported smooth functions that one expects to assign to $d\beta$.
We can introduce polar coordinates $z= r e^{i\theta}$. Then 
\begin{align}
\frac{dz}{z}=\frac{dr}{r}+id\theta=d\ln r +id\theta.
\end{align}
It follows that
\begin{align}
 d\left( \frac{dz}{z}\right)[f]&=\int_{\mathbb{C}} \frac{dz}{z}\wedge df \nonumber \\ &=\int_{\mathbb{C}}dr\wedge \frac{1}{r}\frac{\partial f}{\partial \theta } d\theta +i\int_{\mathbb{C}}d\theta\wedge \frac{\partial f}{\partial r}dr \nonumber \\
&=-i\int_{0}^{+\infty}\int_{0}^{2\pi}\frac{\partial f}{\partial r}drd\theta=2\pi i  f(0),
\end{align}
where in going from the first line to the second we used the fact that $f$ is smooth (hence single-valued) and that $dr\wedge d\theta$ defines the standard orientation of the plane. Since this is valid for any compactly supported $f$, we have that, as a current,
\begin{align}
-d\left(\frac{dz}{z}\right)=-2\pi i \delta(x) dx^1\wedge dx^2,
\end{align}  
where $\delta$ denotes the Dirac delta distribution over $\mathbb{C}$ (a $0$-current). We would then be tempted, using the fact that the exterior derivative commutes with pullbacks, to write the curvature $\Omega$, restricted to $U_\alpha$---meaning that, as a $2$-current, it will act only on smooth differential $\left(m-2\right)$-forms with compact support contained in $U_\alpha$), as
\begin{align}
\Omega\Big|_{U_\alpha}=\Omega_{\alpha}=d\omega_\alpha=-dd\ln f_{\alpha}=-2\pi i f_\alpha^* \left(\delta(x)dx^1\wedge dx^2\right).
\label{eq: local curvature}
\end{align}
For the above equation make sense, we need to, if possible at all, give an appropriate meaning to the pullback $f_\alpha^* \left(\delta(x)dx^1\wedge dx^2\right)$ as a $2$-current over $U_{\alpha}$. We first note that since the support of the $2$-current $\delta(x)dx^1\wedge dx^2$ is the origin of $\mathbb{R}^2$, we see that the support of its pullback, if well-defined, should be contained in $f_{\alpha}^{-1}\left(\{0\}\right)$ which is the intersection of the zero locus of $s$, denoted $Z=\{x\in M: s(x)=0\}$, with $U_{\alpha}$. Provided the tangent map $df_{\alpha}$ is surjective over $Z\cap U_{\alpha}$, which is equivalent to $f_\alpha^* \left(dx^1\wedge dx^2\right)\neq 0$ everywhere on $Z\cap U_\alpha$, we have that $f_{\alpha}^{-1}\left(\{0\}\right)$ is a submanifold of $U_{\alpha}$ by the regular value theorem. If this happens for all the $U_{\alpha}$'s in the open cover of $M$, then $Z$ is a submanifold of $M$. These local conditions mean that the image of $s$, $s(M)\subset \mathscr{L}$, is \emph{transverse} to the zero section of $\mathscr{L}$, implying that the zero locus of $s$, $Z=\{x\in M: s(x)=0\}$, is a submanifold of $M$ of dimension $m-2$. A section $s$ with that property is referred to as a \emph{generic section} and, from now on, we will assume this is the case. As a submanifold, $Z$ has a natural orientation which we proceed to describe. Without loss of generality, we may assume that $U_\alpha$ is a coordinate neighbourhood, and over $U_\alpha$ we describe $M$ by local coordinates $x^1,\dots, x^{m}$ consistent with the orientation of $M$. Over $U_\alpha$ we can describe $\mathscr{L}$ as $U_\alpha\times \mathbb{C}$ by writing any element of $\mathscr{L}$ as $e_\alpha(x) a$, with $x\in U_\alpha$ and $a\in\mathbb{C}$. Here $a$ is the fiber coordinate. Note that, by definition of $f_{\alpha}$ we have that the composition $a\circ s(x)= f_{\alpha}(x)$ gives rise to the local map $f_\alpha:U_\alpha\to \mathbb{C}$. The total space $\mathscr{L}$ has also a natural orientation which is locally determined by declaring $x^1,\dots, x^m, \text{Re}(a), \text{Im}(a)$ is a positive coordinate system. Changing amongst coordinate systems of this type gives rise to a positive Jacobian so this defines an orientation on $\mathscr{L}$. The zero section of $\mathscr{L}$ is locally determined by the equation $a=0$ and hence it is a smooth submanifold of $\mathscr{L}$ locally determined by the coordinates $x^1,\dots, x^m$. We orient it by simply declaring that $x^1,\dots, x^m$ is positively oriented. We can then orient $Z$ using the following argument. By shrinking $U_\alpha$ if necessary, we may use positively oriented coordinates $y^1,\dots, y^{m-2},y^{m-1}=\text{Re}(f_{\alpha}),y^{m}=\text{Im}(f_{\alpha})$ which have the property that $Z\cap U_\alpha=\{y^{m-1}=y^{m}=0\}$. The coordinates $y^1,\dots, y^{m-2}$ define local coordinates on $Z$ and we declare them to be positively oriented. The reason why this is well-defined is because $d\text{Re}(f_{\alpha}), d\text{Im}(f_{\alpha})$ form a local oriented frame for the co-normal bundle of $Z$, which is the pullback of the co-normal bundle of the zero section, an oriented bundle whose orientation is such that $d\text{Re}(a),d\text{Im}(a)$ is positive. Using this arguments we can finally determine the action of $d \left(\frac{d f_\alpha}{f_\alpha}\right)$ on a $\left(m-2\right)$-form $\eta$ with compact support contained in $U_\alpha$ and give meaning to the right-hand side of Eq.~\eqref{eq: local curvature}. Just like in $\mathbb{C}$, here integration by parts and the wedge product formula tells us that we should define
\begin{align}
d \left(\frac{d f_\alpha}{f_\alpha}\right)[\eta]=\frac{d f_\alpha}{f_\alpha}[d\eta]=\int_{U_\alpha} \frac{d f_\alpha}{f_\alpha}\wedge d\eta.  
\end{align}
Since $U_\alpha$ is, by construction, conveniently described by coordinates $y^1,\dots, y^{m-2},y^{m-1}=\text{Re}(f_{\alpha}),y^{m}=\text{Im}(f_{\alpha})$ we may use them also to describe $\eta$. The coordinates provide a map to an open set in $\mathbb{R}^m$. We can extend $\eta$, now interpreted as a differential form in the image of $U_\alpha$ in $\mathbb{R}^m$, by zero away from the image of its support in $\mathbb{R}^m$, and we can therefore think of the above integral as an ordinary integral in $\mathbb{R}^m$, taking into account that $dy^1\wedge \dots \wedge dy^m$ is in the positive orientation.  Additionally, just like in the plane, we will use polar coordinates to describe $f_\alpha=y^{m-1}+i y^{m}=r e^{i\theta}$, replacing the last two coordinates---i.e. we will use positively oriented coordinates $y^1,\dots , y^{m-2},r,\theta$. We will compactly denote $y=(y^1,\dots,y^{m-2})$ and $(y,f_\alpha)$ will represent a point in $U_\alpha$. We can then write
\begin{align}
\frac{df_{\alpha}}{f_{\alpha}}=\frac{dr}{r} +id\theta,
\end{align}
and we may write 
\begin{align}
\eta &=b_{1\dots m-2}(y,f_\alpha) dy^{i_1}\wedge \dots\wedge dy^{i_{m-2}}  \nonumber \\
&+ \sum_{I} b_{r I}(y,f_\alpha) dr\wedge dy^{i_1}\wedge \dots \wedge dy^{i_{m-3}}\nonumber \\
&+\sum_{I}b_{\theta I}(y,f_\alpha) d\theta\wedge dy^{i_1}\wedge \dots \wedge dy^{i_{m-3}}\nonumber \\
&+\sum_{J}b_{r\theta J}(y,f_\alpha)dr\wedge d\theta\wedge dy^{j_1}\wedge \dots \wedge dy^{j_{m-4}},
\end{align}
where we use multi-index notation for simplicity, i.e., $I=(i_1,\dots, i_{m-3})$ and $J=(j_1,\dots,j_{m-4})$ with the $i$ and $j$'s taking values in $1,\dots, m-2$ and where the $b$'s are smooth functions. Now
\begin{widetext}
\begin{align}
d\eta &=\frac{\partial b_{1\dots m-2}}{\partial r}(y,f_\alpha) dr\wedge dy^{i_1}\wedge \dots \wedge dy^{i_{m-2}} +\frac{\partial b_{1\dots m-2}}{\partial \theta}(y,f_\alpha) d\theta\wedge dy^{i_1}\wedge \dots \wedge dy^{i_{m-2}}  \nonumber \\
&+ \sum_{I}\left( \frac{\partial b_{r I}}{\partial \theta}(y,f_\alpha) d\theta\wedge dr\wedge dy^{i_1}\wedge \dots \wedge dy^{i_{m-3}}+\sum_{i}\frac{\partial b_{r I}}{\partial y^i}(y,f_\alpha) dy^i\wedge dr\wedge dy^{i_1}\wedge \dots \wedge dy^{i_{m-3}}\right)\nonumber \\
&+\sum_{I}\left( \frac{\partial b_{\theta I}}{\partial r}(y,f_\alpha) dr\wedge d\theta\wedge dy^{i_1}\wedge \dots \wedge dy^{i_{m-3}}+\sum_{i}\frac{\partial b_{\theta I}}{\partial y^i}(y,f_\alpha) dy^i\wedge d\theta\wedge dy^{i_1}\wedge \dots \wedge dy^{i_{m-3}}\right)\nonumber \\
&+\sum_{J}\sum_{i}\frac{\partial b_{r\theta J}}{\partial y^i}(y,r\theta)dy^i\wedge dr\wedge d\theta\wedge dy^{j_1}\wedge \dots \wedge dy^{j_{m-4}}.
\end{align}
Due to anti-symmetry of the wedge product, we obtain
\begin{align}
\frac{df_{\alpha}}{f_{\alpha}}\wedge d\eta &=\left( \frac{dr}{r}+id\theta\right)\wedge d\eta \nonumber\\
&=i\frac{\partial b_{1\dots m-2}}{\partial r}(y,f_\alpha) d\theta\wedge dr\wedge dy^{i_1}\wedge \dots \wedge dy^{i_{m-2}} +\frac{1}{r}\frac{\partial b_{1\dots m-2}}{\partial \theta}(y,f_\alpha) dr\wedge d\theta\wedge dy^{i_1}\wedge \dots \wedge dy^{i_{m-2}}  \nonumber \\
&+ i\sum_{I}\sum_{i}\frac{\partial b_{r I}}{\partial y^i}(y,f_\alpha) d\theta\wedge dy^i\wedge dr\wedge dy^{i_1}\wedge \dots \wedge dy^{i_{m-3}}\nonumber \\
&+\sum_{I}\sum_{i}\frac{1}{r}\frac{\partial b_{\theta I}}{\partial y^i}(y,f_\alpha) dr\wedge dy^i\wedge d\theta\wedge dy^{i_1}\wedge \dots \wedge dy^{i_{m-3}}.
\end{align}
\end{widetext}
Due to the compact support assumption on $\eta$, all terms containing derivatives $\frac{\partial}{\partial y^i}$ will give zero by applying the Barrow rule. The ones containing $\frac{\partial}{\partial \theta}$ will also give zero since the differential form is smooth. The result is
\begin{align}
&\int_{U_\alpha} \frac{df_{\alpha}}{f_{\alpha}}\wedge d\eta
\nonumber \\
&=\int_{\mathbb{R}^m}i\frac{\partial b_{1\dots m-2}}{\partial r}(y,f_\alpha) d\theta\wedge dr\wedge dy^{i_1}\wedge \dots \wedge dy^{i_{m-2}} \nonumber \\
&=-\int_{\mathbb{R}^m}i\frac{\partial b_{1\dots m-2}}{\partial r}(y,f_\alpha) dy^1\dots dy^{m-2} dr d\theta \nonumber \\
&= -\int_{0}^{+\infty}\int_{0}^{2\pi}\int_{\mathbb{R}^{m-2}}i\frac{\partial b_{1\dots m-2}}{\partial r}(y,f_\alpha) dy^1\dots dy^{m-2} dr d\theta, \nonumber \\
&=2\pi i\int_{\mathbb{R}^{m}} b_{1\dots m-2}(y,0) dy^1\dots dy^{m-2}
\end{align}
where we used that $dy^{i_1}\wedge \dots \wedge dy^{i_{m-2}}\wedge dr\wedge d\theta$ is in the positive orientation. Observe then that if $j:Z\to M$ is the inclusion, then
\begin{align}
j^*\eta= b_{1\dots m-2}(y,0)dy^1\wedge \dots \wedge dy^{m-2}.
\end{align}
We conclude then that
\begin{align}
\Omega_{\alpha}[\eta]&=-d \left(\frac{d f_\alpha}{f_\alpha}\right)[\eta]=-\left(\frac{d f_\alpha}{f_\alpha}\right)[d\eta]=-2\pi i\int_{f_{\alpha}^{-1}(0)} j^*\eta \nonumber \\
&=-2\pi i\int_{U_\alpha\cap Z}j^*\eta=-2\pi i\int_{Z}j^*\eta,
\label{eq: local curvature 2-current}
\end{align}
where we used the fact that the support of $\eta$ is contained in $U_\alpha$. This is consistent with a naive interpretation of the right-hand side of Eq.~\eqref{eq: local curvature}, in the case when $s$ is generic and $Z$ is a submanifold. Indeed, in the coordinates $y^1,\dots, y^{m-2},y^{m-1}=\text{Re}(f_{\alpha}), y^{m}=\text{Im}(f_{\alpha})$, Eq.~\eqref{eq: local curvature} would give
\begin{align}
\Omega_{\alpha} &=-2\pi i f_\alpha^* \left(\delta(x)dx^1\wedge dx^2\right) \nonumber \\ 
&=-2\pi i \delta(y^{m-1})\delta(y^{m})dy^{m-1}\wedge dy^{m},
\end{align}
and the standard rules for manipulation of $\delta$ symbols produce the same result as above. We point out that special care has to be taken in the case $m=2$, because then $m-2=0$, $Z$ is a collection of isolated points and $y^{m-1}=y^{1}=\text{Re}(f_{\alpha}),y^{m-2}=y^2=\text{Im}(f_{\alpha})$ do not form a positively oriented coordinate system for $U_\alpha$ if $f_{\alpha}$, which here can be seen as a map from a plane to a plane, is not orientation preserving over $Z$. One can remedy this by assigning a sign $\sigma_{p}\in \{\pm\}$, depending on whether $f_{\alpha}$ is orientation preserving or reserving at $p\in Z$. In that case, a $(m-2)$-form is a function, say $g$, and
\begin{align}
\Omega_{\alpha}[g]=\sum_{p\in Z}\sigma_p \; g(p).
\label{eq: local curvature 2-current dim 2}
\end{align}
Since the results of Eq.~\eqref{eq: local curvature 2-current} and, for $m=2$, Eq.~\eqref{eq: local curvature 2-current dim 2} hold for all neighbourhoods $U_{\alpha}$ in the open covering, we see that the curvature behaves like a Dirac delta concentrated on the zero locus of $s$, because it forces the $f_{\alpha}$'s, which are the local components of $s$ in the trivializations $e_{\alpha}$'s, to zero. We can show that the result is global as follows. By taking a partition of unity $\{\phi_\alpha\}$ subordinate to the cover $\{U_{\alpha}\}$, we may write a $\left(d-2\right)$-form $\eta$ with compact support as $\sum_{\alpha} \phi_{\alpha} \eta$. Now the support of $\phi_{\alpha}\eta$ is contained on a compact subset of $U_\alpha$, for each $\alpha$. We may consistently define
\begin{align}
\Omega[\eta]&:=\int_{M}\Omega\wedge \eta=\int_{M}\Omega\wedge \left(\sum_\alpha \phi_\alpha \eta\right) \nonumber \\ &=\sum_{\alpha}\int_{U_{\alpha}}\Omega_{\alpha}\wedge \left(\phi_{\alpha}\eta\right)=\sum_{\alpha}\Omega_{\alpha}[\phi_{\alpha}\eta],
\end{align}
where the sum on the right-hand side is finite because, by assumption, $\eta$ is of compact support. Finally, from the above, we know that $\Omega_{\alpha}[\phi_{\alpha}\eta]=\int_{Z}j^*\left(\phi_\alpha \eta\right)$ and hence
\begin{align}
\Omega[\eta] &=\sum_{\alpha}\Omega_{\alpha}[\phi_{\alpha}\eta]=-2\pi i\sum_{\alpha}\int_{Z}j^*\left(\phi_\alpha \eta\right) \nonumber \\ &=-2\pi i\int_{Z}j^*\left(\sum_{\alpha}\phi_\alpha \eta\right)=-2\pi i\int_{Z}j^*\eta.
\end{align}

Accordingly, the curvature $\Omega$ will be a $2$-current which, under the assumption that $Z=\{x\in M: s(x)=0\}$ is a submanifold, just assigns the number
\begin{align}
\Omega [\eta]:=\int_{M}\Omega\wedge \eta=-2\pi i \int_{Z}j^*\eta,
\end{align}
to any $\eta$ a smooth differential $\left(m -2\right)$-form with compact support, where $j:Z\hookrightarrow M$ is the inclusion. This result is consistent with the well-known fact that the first Chern class of a line bundle is \emph{Poincaré dual} to the zero locus of a generic section. In particular, when $M$ is a compact $2$-manifold, we get the standard result that the first Chern number is just counting (algebraically) the zeros of a (generic) section of $\mathscr{L}$, since in that case
\begin{align}
\frac{i}{2\pi}\Omega [1]=\int_{M}\frac{i\Omega}{2\pi}=\sum_{p\in Z}\sigma_p,
\label{eq: 1st Chern number as sum over zeros}
\end{align}
where $\sigma_p\in\{\pm 1\}$ depending on whether the local map (defined using a local trivialization of $\mathscr{L}$) $s: U_{p}\subset \mathbb{R}^2 \to \mathbb{C}\cong \mathbb{R}^2$ is orientation preserving or reversing at $p$, where $U_p\subset M$ is a local neighbourhood containing the zero $p\in Z$.

\section{Spins in magnetic fields---general construction}
\label{sec: spins in magnetic fields---general construction}
We consider the Hamiltonian of a particle of spin $s$ interacting with a magnetic field via 
\begin{align}
H(\vec{B})=\vec{B}\cdot \vec{S},
\label{eq: hamiltonian for spin in magnetic field}
\end{align}
where $\vec{B}=(B_1,B_2,B_3)\in\mathbb{R}^3$ is the magnetic field and $\vec{S}=(S_1,S_2,S_3)$ are the generators of $\text{SU}(2)$ determining a spin $s$ irreducible representation, whose dimension is $2s+1$.
The eigenenergies of the above Hamiltonian are 
\begin{align}
|\vec{B}|m, \text{ with } m\in \{-s,\dots, s\}.
\end{align}

In the main text, we analyzed in detail the case of $s = 1/2$. Here we discuss the general cases of $s \ge 1/2$ and outline how, using a holomorphic approach, one can build anti-holomorphic and holomorphic, respectively for $m>0$ and $m<0$, sections of $\mathscr{L}_m$. These sections, described by homogeneous polynomials in two complex variables $f(z_0,z_1)$ (or their complex conjugates) of degree $2|m|$, have exactly $2|m|$ zeros in the Bloch sphere. The curvature of the associated singular connection, as we see, is $2\pi i\; \text{sgn}(m)$ times a Dirac delta localized in the positions of these zeros on the Bloch sphere, producing the well-known result that $\mathscr{L}_m$, when restricted to a sphere enclosing the origin of radius $|\vec{B}|\neq 0$, has first Chern number $-2m$.

We consider $\vec{B}\neq 0$, and first focus on the line bundle $\mathscr{L}_{-s}$ over parameter space determined by the lowest energy eigenvalue $-s|\vec{B}|$, i.e., for $m=-s$. While a specific local expression for the eigenstates of Eq.~\eqref{eq: hamiltonian for spin in magnetic field} might seem hard at first, one can workout all the details with the following representation theoretical argument. For fixed $\vec{B}$ there exists matrix in this representation of $\text{SU}(2)$ which diagonalizes $\vec{B}\cdot \vec{S}$, and this is precisely the image under the representation map of the one diagonalizing this Hamiltonian for $s=1/2$, i.e., in the fundamental representation. In the fundamental representation $\vec{S}=(1/2)(\sigma_1,\sigma_2,\sigma_3)$ where $\sigma_i$'s are the Pauli matrices. In that case we may write
\begin{align}
H(\vec{B}) &=\frac{1}{2}\left[\begin{array}{cc}
B_3 & B_1-iB_2\\
B_1 + iB_2 & -B_3
\end{array}\right] \nonumber \\
&=\frac{1}{2} U\left[\begin{array}{cc}
|\vec{B}| &  0\\
   0  &  -|\vec{B}|
\end{array}\right]U^{\dagger}\nonumber \\
&=\frac{|\vec{B}|}{2} U\left[\begin{array}{cc}
 1 &  0\\
   0  &  -1
\end{array}\right]U^{\dagger}=\frac{|\vec{B}|}{2}\vec{n}\cdot \vec{S},
\end{align}
where $U\in\text{SU}(2)$ diagonalizes our $2\times 2$ problem and $\vec{n}=\vec{B}/|\vec{B}|$. Notice how the particular value of $|\vec{B}|$ is irrelevant of the diagonalization of $H(\vec{B})$. Observe that if $U$ is one such matrix then $U g$, where $g$ is a diagonal $\text{SU}(2)$ matrix, is another valid choice for the diagonalization of $H(\vec{B})$. In other words, effectively, the Hamiltonian is parameterized by the quotient $\text{SU}(2)/\text{U}(1)\cong S^2$, i.e., the Bloch sphere described by $\vec{n}=\frac{\vec{B}}{|\vec{B}|}$. In particular, if we write the complex coordinate 
\begin{align}
z=\frac{B^1+iB^2}{|\vec{B}|+B^3},    
\end{align}
corresponding to stereographic projection with respect to the south pole of the sphere, we can write a parametrization of $U$, valid for $\vec{B}\neq (0,0,-|\vec{B}|)$, given by
\begin{align}
U=\frac{1}{\left(1+|z|^2\right)^{\frac{1}{2}}}\left[\begin{array}{cc}
1 & -\bar{z}\\
z & 1
\end{array}
\right]    
\end{align}
Furthemore, writing the complex coordinate 
\begin{align}
w=\frac{1}{z}=\frac{B^1-iB^2}{|\vec{B}|-B^3},    
\end{align}
corresponding to stereographic projection with respect to the north pole of the sphere, we can write a parametrization of $U$, valid for $\vec{B}\neq (0,0,|\vec{B}|)$ by
\begin{align}
U'=\frac{1}{\left(1+|w|^2\right)^{\frac{1}{2}}}\left[\begin{array}{cc}
w & -1\\
1 & \bar{w}
\end{array}
\right].
\end{align}
The relation between the two gauges in region of parameter space where both are valid, $\vec{B}\neq (0,0,\pm |\vec{B}|)$, is given by
\begin{align}
U=U'\left[\begin{array}{cc}
\frac{z}{|z|} &  0 \\
0 & \frac{\bar{z}}{|z|}
\end{array}\right].
\end{align}
From this discussion, it will now become clear what the bundles $\mathscr{L}_{m}$, $m\in\{-s,\dots,s\}$ are, because under the representation map the matrix $\text{diag}(\frac{z}{|z|},\frac{\bar{z}}{|z|})$ is mapped to
\begin{align}
\text{diag}\left[\left(\frac{z}{|z|}\right)^{2m}:m=-2s,\dots, 2s\right].
\end{align}
Since we may view a matrix $U\in\text{SU}(2)$ as a point on the three dimensional sphere labeled by $(z_0,z_1)$ in the first column of $U$, it follows that the bundle $\mathscr{L}_{m}$ is determined by the set of pairs $((z_0,z_1), v)\in S^3\times \mathbb{C}$ under the equivalence relation of $\text{U}(1)$-gauge transformations:
\begin{align}
 ((z_0,z_1), v)\sim ((e^{i\phi} z_0, e^{i\phi} z_1), e^{-2m i \phi} v). 
\end{align}
The reason for this can be seen by looking at what happens for $s=1/2$ and $m=1/2$, where we have that the pairs $((1,z),v)$ and $((e^{i\phi},e^{i\phi}z) ,e^{-i\phi}v)$ determine the same eigenstate:
\begin{align}
\frac{1}{\left(1+|z|^2\right)^{\frac{1}{2}}}\left[\begin{array}{cc}
1\\
z
\end{array}\right]v=\frac{1}{\left(1+|z|^2\right)^{\frac{1}{2}}}\left[\begin{array}{cc}
e^{i\phi}\\
ze^{i\phi}
\end{array}\right]e^{-i\phi}v.
\end{align}
With this description at hand, we may construct non-trivial sections of the bundle $\mathscr{L}_{-s}$ in a very convenient way. We first realize that $\mathscr{L}_{-s}$ may be understood in the holomorphic setting by replacing $S^3$ by $\mathbb{C}^2-\{0\}$ and $\text{U}(1)$ by $\mathbb{C}^*=\mathbb{C}-\{0\}$ and considering $S^2$ as a complex manifold, namely, considering it as $\mathbb{C}P^1=\mathbb{C}^2-\{0\}/\mathbb{C}^*$. Then we are looking at pairs $((z_0,z_1),v)$ under the more general equivalence relation
\begin{align}
((z_0,z_1),v)\sim ((z_0\lambda,z_1\lambda), \lambda^{2s} v),    
\end{align}
for $\lambda\in \mathbb{C}^*$. A holomorphic section of $\mathscr{L}_{-s}$ is then determined by considering a holomorphic function $f(z_0,z_1)$ and demanding that $((z_0,z_1),f(z_0,z_1))$ determines a well-defined equivalence class , i.e.,
\begin{align}
f(\lambda z_0,\lambda z_1)=\lambda^{2s} f(z_0,z_1),
\end{align}
i.e., $f$ must be a homogeneous polynomial of degree $2s$. We denote the associated section by $s_f$. Such polynomial is a linear combination of $z_{0}^{2s-m}z_1^{m}$, $m\in\{0,\dots,2s\}$,
\begin{align}
f(z_0,z_1)=\sum_{m=0}^{2s}a_m z_0^{2s-m}z_1^m.
\end{align}
A homogeneous polynomial of degree $2s$ will have exactly $2s$ zeros in $\mathbb{C}P^1$. To see this, write,
\begin{align}
f(z_0,z_1)=z_0^{2s}f(1,z),
\end{align}
with $z=z_1/z_0$ and $f(1,z)=\sum_{m=0}^{2s}a_m z^m$. Provided $a_{2s}\neq 0$, $f(1,z)$ is a degree $2s$ ordinary polynomial in $z$ it has exactly $2s$ roots. This covers $z\neq \infty$, i.e., $z_0\neq 0$. We can also write
\begin{align}
f(z_0,z_1)=z_1^{2s}f(w,1),
\end{align}
where now $f(w,1)=\sum_{m=0}^{2s}a_{m}w^{2s-m}$. For the section associated to $f$ to have zeros at $z=\infty$, we need $f(0,1)=0$ implying $a_{2s}=0$ which contradicts our initial hypothesis. Now supposes $a_{2s}=a_{2s-1}=...=a_{m_0+1}=0$, $m_0\leq 2s$. Then $f(1,z)$ is a polynomial of degree $m_0$ and it follows that it has $m_0$ roots. Now near infinity we can write $z_1^{2s}f(w,1)$. Observe that $f(w,1)=\sum_{m=0}^{m_0}a_{m}w^{2s-m}=a_{0}w^{2s}+\dots +a_{m_0}w^{2s-m_0}=w^{2s-m_0}\left(a_0 w^{m_0}+\dots a_{m_0}\right)$ meaning that $f$ has a zero of order $2s-m_0$ at $w=0$, i.e., at $z=\infty$. We then conclude that $f$ has $2s$ zeros and so does the associated section. Since $m_0$ was arbitrary, this completes the proof. 

Now we take a collection of $2s$ independent complex numbers $z_0,\dots, z_{2s}$, non coincident (hence, generic), and take $f(z_0,z_1)=Cz_0^{2s}\prod_{j=1}^{2s}(z-z_j)$, with $z=z_1/z_0$, for some constant $C\neq 0$. The associated section will have zeros away from $z=\infty$, precisely at the $z_j$'s. We can define a singular connection by declaring $\nabla s_f\equiv 0$. The associated curvature two-form is given by 
\begin{align}
F^{\text{sing}}=-2\pi i \sum_{j=1}^{2s} \delta^2(z-z_j)\frac{ idz\wedge d\bar{z}}{2},
\end{align}
where $\delta^2(z)\frac{ idz\wedge d\bar{z}}{2}$ is such that for a test function $g(z)$ we have $\int_{\mathbb{C}} g(z)\delta^2(z)\frac{ idz\wedge d\bar{z}}{2}=g(0)$. The reason for weights of each $\delta$ function being $+1$ for each $z_j$ is related to the fact that locally the section determines a holomorphic map to the complex plane and these are always orientation preserving. Note that we can represent the $z_j$'s in the Bloch sphere through the formula
\begin{align}
\vec{n}_j=\frac{1}{\left(1+|z_j|^2\right)}\left[ 1 \ \bar{z}_j
\right]\vec{\sigma}\left[\begin{array}{cc}
     1  \\
     z_j 
\end{array}\right],
\end{align}
and the above formula says that $F^{\text{sing}}$ is concentrated at the values $\vec{B}=|\vec{B}| \vec{n}_j$, $j=1,\dots, 2s$. Observe that the first Chern number, computed by restricting on a sphere of constant $|\vec{B}|$, is then immediately given by counting the zeros, i.e., it is equal to $2s$.

We point out that similar construction can be done for the bundles $\mathscr{L}_{|m|}$ for $m\leq 0$. There the sections will consist of homogeneous polynomials of degree $2m$. Choosing now $2m$ different complex numbers $z_j$'s, determining a section up to scale, we obtain the analogous formula for the curvature
\begin{align}
F^{\text{sing}}=-2\pi i \sum_{j=1}^{2|m|} \delta^2(z-z_j)\frac{ idz\wedge d\bar{z}}{2}.
\end{align}

For $m>0$, however, the most convenient thing to do is to use anti-holomorphic sections of the form $\overline{f(z_0,z_1)}$, for some (generic) homogeneous polynomial of degree $2m$---the reason being that the line bundle $\mathscr{L}_{m}$, for $m>0$, has no holomorphic sections. In that case, because the associated sections can be locally described by anti-holomorphic maps of the plane to itself and these are always orientation reversing, the analogous formula is
\begin{align}
F^{\text{sing}}=2\pi i \sum_{j=1}^{2m} \delta^2(z-z_j)\frac{ idz\wedge d\bar{z}}{2}.
\end{align}

As a consequence, one recovers the well-known fact that the first Chern number of $\mathscr{L}_{m}$ is $-2m$. Indeed, if we had used the Berry connection, we would have obtained the curvature two-form
\begin{align}
F=i m \; \vec{n}\cdot \left( d\vec{n}\times d\vec{n}\right),   
\end{align}
where $\vec{n}\cdot \left( d\vec{n}\times d\vec{n}\right)$ is the standard area form on the sphere with area $4\pi$, producing exactly the same result, equal to $-2m$, for the first Chern number. 

\bibliography{bib.bib}
\end{document}